\documentstyle[12pt,subeqnarray,psfig]{article}
\begin{document}
%
%
\newenvironment{lefteqnarray}{\arraycolsep=0pt\begin{eqnarray}}
{\end{eqnarray}\protect\aftergroup\ignorespaces}
\newenvironment{lefteqnarray*}{\arraycolsep=0pt\begin{eqnarray*}}
{\end{eqnarray*}\protect\aftergroup\ignorespaces}
\newenvironment{leftsubeqnarray}{\arraycolsep=0pt\begin{subeqnarray}}
{\end{subeqnarray}\protect\aftergroup\ignorespaces}
\newcommand{\appleq}{\stackrel{<}{\sim}}
\newcommand{\appgeq}{\stackrel{>}{\sim}}
\newcommand{\arcsinh}{\mathop{\rm arcsinh}\nolimits}
\newcommand{\arctg}{\mathop{\rm arctg}\nolimits}
\newcommand{\diff}{{\rm\,d}}
\newcommand{\displayfrac}[2]{\frac{\displaystyle #1}{\displaystyle #2}}
\newcommand{\Erfc}{\mathop{\rm Erfc}\nolimits}
\newcommand{\Int}{\mathop{\rm Int}\nolimits}
\newcommand{\Nint}{\mathop{\rm Nint}\nolimits}
\newcommand{\pprime}{{\prime\prime}}

\title{Virialization of matter overdensities within 
       dark energy subsystems: special cases}
\author{{R. Caimmi\footnote{
{\it Astronomy Department, Padua Univ., Vicolo Osservatorio 2,
I-35122 Padova, Italy} 
email: caimmi@pd.astro.it
}
\phantom{agga}}}

\maketitle
\begin{quotation}
\section*{}
\begin{Large}
\begin{center}

Abstract

\end{center}
\end{Large}
\begin{small}


The virialization of matter overdensities within
dark energy subsystems is considered under a
number of restrictive assumptions, namely (i)
spherical-symmetric density profiles, (ii)
time-independent quintessence equation of state
parameter, $w$, and (iii) nothing but gravitational
interaction between dark energy scalar field and
matter.   In addition, the quintessence subsystem
is conceived as made of ``particles'' whose mutual
interaction has intensity equal to $G(1+3w)$ and
scales as the inverse square of their distance.
The related expression of the self and tidal
potential energy and formulation of the virial
theorem for subsystems, are found to be consistent
with their matter counterparts, passing from 
$-1\le w<-1/3$ to $w=0$.   In the special case
of fully clustered quintessence, energy
conservation is assumed with regard to either
the whole system (global energy conservation),
or to the matter subsystem within the tidal
potential induced by the quintessence subsystem
(partial energy conservation).   Further
investigation is devoted to a few special cases,
namely a limiting situation, $w=-1/3$, and three
lower values, $w=-1/2, -2/3, -1$, where the last
one mimics the effect of a cosmological constant.
The special case of fully clustered (i.e.
collapsing together with the matter) quintessence
is studied in detail, using a similar procedure
as in Maor \& Lahav (2005).   The general case of
partially clustered quintessence is considered in
terms of a degree of quintessence de-clustering,
$\zeta$, $0\le\zeta\le1$, ranging from fully
clustered ($\zeta=0$) to completely de-clustered
($\zeta=1$) quintessence, respectively.   The
special case of unclustered (i.e. remaining
homogeneous) quintessence is also discussed.
The trend exhibited by the fractional (virialization to
turnaround) radius, $\eta$, as a function of
the (i) fractional (quintessence to matter) ``mass'' ratio at
turnaround, $m$, (ii) degree of quintessence
de-clustering, $\zeta$, and (iii) quintessence
equation of state parameter, $w$, is found to
be different from its counterparts reported in
earlier attempts.   In particular, no clear
dichotomy with respect to the limiting
situation of vanishing quintessence, $\eta=1/2$,
is shown when global or partial energy conservation
hold in the special case of fully clustered
quintessence, with $\eta>1/2$ preferred.   The
reasons of the above mentioned discrepancy are
recognized as owing to (i) different formulations
of the virial theorem for subsystems, and (ii)
different descriptions of de-clustered quintessence,
with respect to the reference case of fully
clustered quintessence.

\noindent
{\it keywords - Dark matter; Dark energy; 95.35.+d; 95.36.+x.}

\end{small}
\end{quotation}

%

\section{Introduction}\label{intro}
Recent observations from anisotropies in the cosmic microwave
background, large-scale structure surveys, Hubble parameter
determinations, and Type Ia supernova results, allow narrow 
ranges for the values of cosmological quantities (e.g.,
Sievers et al. 2003; Rubig\~no-Martin et al. 2003;
Spergel et al. 2003).   The related ``corcondance''
cosmological model is consistent with a bottom-up picture
(hierarchical clustering) of dark matter haloes, where
smaller systems formed first from initial density perturbations
and then merged with each other to become larger systems, or
were tidally disrupted and accreted from successively formed,
more massive neighbours.   A main feature of the concordance
model is that the dominant (about 70\%) contribution to the
present-day energy budget is a component with equation of
state, $p=wc^2\rho$, called dark energy, where $p$ is the
pressure, $\rho$ the energy density, $c$ the speed of the
light, and $w$ a dimensionless parameter which is, in general,
time-dependent.

An obvious candidate, a scalar field, has to be light enough
to vary slowly during a Hubble time, in such a way its potential
energy can drive an accelerated expansion, just like during
inflation.   The varying field equation of state can then be
tuned to lie in the observed range, and the related scalar
field is sometimes called ``quintessence'', which may be
conceived as a fifth component of the cosmic fluid after
ordinary matter (baryons, leptons, and radiation) and
nonbaryonic dark matter.   The scalar field density fraction,
$\Omega_q$, can be made to decrease rapidly in the past, so
as to pass easily the lensing constraints and to avoid
discrepancies in the primordial nucleosynthesis abundances.
For further details, see e.g., Amendola (2000).

The parameter of the equation of state, $w$, depends on how
the scalar field is slowly rolling down its potential.   In
the limit of a completely flat potential, $w=-1$, the
quintessence behaves as a cosmological constant, $\Lambda>0$
(Wetterich 1988; Peebles \& Ratra 1988; Ratra \& Peebles
1988).   A subclass of quintessence models with constant $w$
was proposed by Caldwell et al. (1998a,b) within the range,
$-1\le w<-1/3$.   In fact, $w>-1/3$ corresponds to decelerate
expansion while $w<-1$ implies a phisical interpretation
which is not still clear up today (e.g., Maor \& Lahav 2005,
hereafter quoted as ML05).   Data on large-scale structures
suggest a preferential range, $-1\le w\le-0.6$ (e.g.,
Weinberg \& Kamionkowski 2003).   Many other forms have
been proposed for the shape of the potential of the scalar
field, leading to an equation of state parameter that is
dependent on the scale factor (see Peebles \& Ratra 2003,
for a review).   For further details see e.g., Percival
(2005).

Dark energy affects not only the expansion rate of the
background and the distance-redshift relation, but also
the growth of structure.   The formation rate of haloes,
their evolution and their final characteristics are modified.
Dark energy is therefore expected to have an impact on
observables such as cluster number counts and lensing
statistics due to intervening concentrations of mass on
the line of sight of background sources.   For further
details see e.g., Horellou \& Berge (2005).   The assumption
of an Einstein-de Sitter cosmology, $\Omega_m+\Omega_q=1$,
according to the concordance model, allows a much simpler
analysis, but additional assumptions are needed.

First, overdensities are conceived as spherical-symmetric,
as initially done by Gunn \& Gott (1972).   The related top
hat spherical collapse formalism has been proven to be a
powerful tool for understanding the formation and the
evolution of bound systems in the universe.   For reasons
of simplicity, the special case of homogeneous overdensities
is currently considered (e.g., Lahav et al. 1991; Wang \&
Steinhardt 1998; Iliev \& Shapiro 2001; Horellou \& Berge
2005; Percival 2005; ML05).

Second, the quintessence equation of state is restricted to
constant values of the parameter, $w$.   A time-dependent
parameter, $w$, can result from a changing ratio of
quintessence kinetic to potential energy which, in turn, is 
owing to the evolution of the scalar field potential (e.g.,
Caldwell et al. 1998b).   Though time-varying equations of
state have been used in the literature (e.g., Wetterich
1995; Amendola 2000; Battye \& Weller 2003; Mota \& van de
Bruck 2004; Percival 2005), the special case of constant
$w$ makes considerable simplification (e.g., Caldwell et
al. 1998b; Wang \& Steinhardt 1998; Weinberg \& Kamionkowski
2003; Horellou \& Berge 2005; ML05).

Third, any couplings of quintessence to other fields are
supposed to be negligibly small, so that the scalar field
interacts with other matter only gravitationally.   Different
interactions can be taken into consideration (e.g., Wetterich
1995; Amendola 2000) at the price of less simple analysis
involving a large number of parameters.

In this view, the quintessence is conceived as lying between
two limiting cases, namely (i) full clustering i.e. the
scalar field responds to the infall in the same way as matter,
and (ii) unclustering i.e. its sole effect is a tidal potential
acting on the matter overdensity.   The general case of partial
clustering can be taken into consideration, at the expense of a
much more complicated continuity equation for the quintessence
overdensity (e.g., Mota \& van de Bruck 2004; ML05).   Even
though it has been shown that quintessence cannot be perfectly
smooth (e.g., Caldwell et al. 1998a,b), clustering is usually 
assumed to be negligible on scales less than about 100 Mpc
(e.g., Wang \& Steinhardt 1998; Weinberg \& Kamionkowski 2003;
Battye \& Weller 2003; Horellou \& Berge 2005).   It is
therefore common practice to keep the quintessence homogeneous
during the evolution of the system.   The effects of relaxing
this assumption were explored in recent attempts (e.g., Mota
\& van de Bruck 2004; Percival 2005).   The real situation
may safely be expected to lie between two limiting cases,
where the quintessence is clustering together with the matter
and remains homogeneous, respectively.

In spite of the simplifying assumptions mentioned above,
still different authors use different expressions for (i)
the gravitational potential induced by the quintessence,
(ii) the self and the tidal potential energy, and (iii)
the formulation of the virial theorem, which yields different
results.   On the other hand, considerable effort has been
devoted to the fact, that the quintessence is indistinguishable
from a cosmological constant in the special case, $w=-1$ (e.g.,
Wang \& Steinhardt 1998; Mota \& van de Bruck 2004; Horellou
\& Berge 2005; Percival 2005; but see also ML05; Wang 2006),
but little consideration has been taken on the fact, that the
quintessence is indistinguishable from matter in the special
case, $w=0$ (e.g., Caldwell et al. 1988b; Iliev \& Shapiro
2001).   Accordingly, the expression of the gravitational
potential induced by the quintessence subsystem within a
density perturbation, is expected to reduce to its matter
counterpart as $w\to0$.   In fact,
quintessence and matter gravitational potential are sometimes
unified in a single expression, where $w<-1/3$ corresponds to
the former case and $w=0$ to the latter (e.g., ML05).

The current paper is restricted to the virialization of matter
overdensities within quintessence subsystems, under the
simplifying assumptions mentioned above and, in addition, the
boundary condition that the quintessence potential reduces to
the gravitational potential, when the pressure term is suppressed.
The gravitational
potential and potential energy terms for matter and quintessence
subsystem, are expressed in Sect.\,\ref{pote}.   The virial
theorem for 2-component systems is specified in Sect.\,\ref
{virte}.   The special case of fully clustered quintessence,
and its extension to the general case of partially clustered
quintessence, are investigated in Sects.\,\ref{cluq} and \ref
{uncq}, respectively.   The related discussion is performed in
Sect.\,\ref{disc}, and some concluding remarks are reported in
Sect.\,\ref{conc}.   Further investigation on a few special
arguments is developed in the Appendix.

\section{Matter and quintessence potential-energy terms}
\label{pote}

Let us take into consideration homogeneous and
spherical-symmetric overdensities made of both
matter and quintessence.   Let the quintessence
equation of state:
\begin{equation}
\label{eq:ques}
p_q=wc^2\rho_q~~;
\end{equation}
be restricted to constant values of the parameter,
$w$, being $p$ the pressure, $\rho$ the density,
$c$ the light velocity in vacuum, and the index,
$q$, denoting quintessence.   Let any couplings of
quintessence to other fields be negligibly small,
so that the scalar field interacts with other
matter only gravitationally.

The overdensity may be conceived as a
two-component system made of matter and quintessence,
respectively.   The related gravitational potential
is (e.g., Mota \& van de Bruck 2004; ML05)%
\footnote{A different sign convention is adopted here.}:
\begin{equation}
\label{eq:pote}
{\cal V}_u(r)=2\pi G(1+3w_u)\rho_u\left(R^2-\frac
{r^2}3\right)~~;\qquad u=m,q~~;
\end{equation}
where $w_m=0$, $w_q=w$, $R$ is the radius of the
overdense sphere, and the index, $m$, denotes
matter.

With regard to matter component, Eq.\,(\ref{eq:pote})
reduces to the well known expression of the gravitational
potential induced by a homogeneous sphere on interior
points, with the boundary condition that the maximum
value is attained at the centre (e.g., MacMillan 1930,
Chap.\,II, \S\,29).   In fact, given an interior point
placed at a distance, $r$, from the centre, the potential
induced from the sphere of radius, $r$, is ${\cal V}^
{({\rm int})}(r)=(4\pi/3)G\rho r^2$, according to
MacLaurin's theorem, and the potential induced from the
corona of radii, $r$ and $R$, is ${\cal V}^{({\rm ext})}
(r)=2\pi G\rho(R^2-r^2)$, according to Newton's theorem,
and the potential induced by the sphere of radius, $R$,
at the point under consideration, is ${\cal V}(r)={\cal V}^
{({\rm int})}(r)+{\cal V}^{({\rm ext})}(r)$, according
to Eq.\,(\ref{eq:pote}) particularized to the matter
subsystem.   For further details see e.g., Caimmi (2003).
An alternative expression is related to the boundary
condition of a null potential at the centre (e.g.,
Iliev \& Shapiro 2001; Horellou \& Berge 2005).

In general, the gravitational potential may be
conceived as induced from a matter distribution
where any two particles, idealized as mass points,
interact with strenght, $G$, according to Newton's
law, $F_{m_im_j}=Gm_im_j/r_{ij}^2$, and the dependence
of the resulting force on the distance, $\partial
{\cal V}/\partial r\propto r$, deduced from
Eq.\,(\ref{eq:pote}), is owing to the selected
density profile, $\rho={\rm const}$.

In this view, the gravitational potential induced
by the quintessence may also be conceived as
arising from a distribution where any two
``particles'', idealized as ``mass points'',
interact with strenght, $(1+3w)G$, according to
a Newton-like law, $F_{m_{qi}m_{qj}}=(1+3w)Gm_{qi}
m_{qj}/r_{ij}^2$, where $m_q$ is the ``quintessence
mass''.   Then the results related to two-component
matter distributions (e.g., Limber 1959; Brosche et
al. 1983; Caimmi et al. 1984; Caimmi \& Secco 1992)
may be generalized to the case, where a subsystem
is made of quintessence.

The self potential energy reads (e.g., Chandrasekhar
1969, Chap.\,2, \S\,10):
\begin{equation}
\label{eq:spe}
\Omega_u=-\frac12\int_{S_u}\rho_u(x_1,x_2,x_3)
{\cal V}_u(x_1,x_2,x_3)\diff^3S~~;\qquad u=m,q~~;
\end{equation}
where $S_u$ is the volume of $u$ subsystem.

The interaction potential energy, the tidal
potential energy, and the residual potential
energy read (e.g., Caimmi \& Secco 1992):
\begin{lefteqnarray}
\label{eq:ipe}
&& W_{uv}=-\frac12\int_{S_u}\rho_u(x_1,x_2,x_3)
{\cal V}_v(x_1,x_2,x_3)\diff^3S~~; \\
\label{eq:tpe}
&& V_{uv}=\int_{S_u}\rho_u(x_1,x_2,x_3)
\sum_{r=1}^3x_r\frac{\partial{\cal V}_v}{\partial
x_r}\diff^3S~~; \\
\label{eq:rpe}
&& Q_{uv}=V_{uv}-W_{uv}~~;\qquad u=m,q~~;\qquad
v=q,m~~;
\end{lefteqnarray}
respectively.

The combination of Eqs.\,(\ref{eq:pote}) and
(\ref{eq:ipe}) yields:
\begin{equation}
\label{eq:Wsim}
\frac{W_{uv}}{1+3w_v}=\frac{W_{vu}}{1+3w_u}~~;
\qquad u=m,q~~;\qquad v=q,m~~;
\end{equation}
in the special case of subsystems obeying the
same equation of state, $w_u=w_v$, the
interaction potential energy is symmetric
with respect to the exchange of indices (e.g.,
MacMillan 1930, Chap.\,III, \S76; Caimmi \&
Secco 1992; Caimmi 2003).

In the special case of concentric,
spherical-symmetric, homogeneous subsystems,
the gravitational potential is expressed by
Eq.\,(\ref{eq:pote}), and Eqs.\,(\ref{eq:spe}),
(\ref{eq:ipe}), (\ref{eq:tpe}), and (\ref
{eq:Wsim}) reduce to (e.g., Caimmi 2003):
\begin{lefteqnarray}
\label{eq:sos}
&& \Omega_u=-\frac{16}{15}\pi^2(1+3w_u)G
\rho_u^2R_u^5=-\frac35(1+3w_u)\frac{GM_u^2}
{R_u}~~;\qquad u=m,q~~; \\
\label{eq:soii}
&& W_{uv}=-\frac35(1+3w_v)\frac{GM_i^2}
{R_i}\frac m{y^3}\left(\frac54y^2-\frac
14\right)~~;\quad u=m,q~~;\quad v=q,m~~;\quad \\
\label{eq:soti}
&& V_{ij}=-\frac35(1+3w_j)\frac{GM_i^2}
{R_i}\frac m{y^3}~~;
\end{lefteqnarray}
where the ``mass'' of the quintessence
subsystem, $M_q$, and the dimensionless
ratios, $m$ and $y$, are defined as:
\begin{lefteqnarray}
\label{eq:Mq}
&& M_q=\frac{4\pi}3R_q^3\rho_q~~; \\
\label{eq:my}
&& m=\frac{M_j}{M_i}~~;\qquad y=\frac{R_j}{R_i}~~;
\qquad R_i\le R_j~~;
\end{lefteqnarray}
and the indices, $i$ and $j$, denote the inner
and the outer component, respectively.

Using the virial theorem for both subsystems
and the whole system, together with Eqs.\,(\ref
{eq:Wsim}), (\ref{eq:soii}) and (\ref{eq:soti}),
the tidal potential energy, $V_{ji}$, takes the
explicit expression:
\begin{lefteqnarray}
\label{eq:sotjA}
&& V_{ji}=-\frac35\frac{GM_i^2}{R_i}\frac m{y^3}
\left(\frac52y^2-\frac32\right)\nonumber \\
&& \phantom{V_{ji}=}\times\left[1+\frac32w_j\frac
{5y^2-5}{5y^2-3}+\frac32w_i\frac{5y^2-1}{5y^2-3}
\right]~~;
\end{lefteqnarray}
and a formal demonstration is provided in the
next section.
The combination of Eqs.\,(\ref{eq:soti}) and 
(\ref{eq:sotjA}) yields:
\begin{lefteqnarray}
\label{eq:sotjB}
&& V_{ji}=V_{ij}\left[\left(\frac54y^2-\frac54\right)
\right.\nonumber \\
&& \phantom{V_{ji}=V_{ij}}\left.+\left(\frac54y^2-
\frac14\right)\frac{1+3w_i}{1+3w_j}\right]~~;
\end{lefteqnarray}
which makes a relation between tidal potential
energies.

Under the further restriction that the two
subsystems are bounded by a single sphere,
$R_i=R_j=R$ or $y=1$, Eqs.\,(\ref{eq:soii}), 
(\ref{eq:soti}) and (\ref{eq:sotjB}) reduce to:
\begin{lefteqnarray}
\label{eq:Wmq}
&& \frac{W_{mq}}{1+3w_q}=\frac{W_{qm}}{1+3w_m}=
-\frac35\frac{GM_mM_q}R~~; \\
\label{eq:Vmq}
&& \frac{V_{mq}}{1+3w_q}=\frac{V_{qm}}{1+3w_m}=
-\frac35\frac{GM_mM_q}R~~;
\end{lefteqnarray}
and the interaction potential energy coincides
with the tidal potential energy, which is the
limiting case currently used in the literature
(e.g., Lahav et al. 1991; Wang \& Steinhardt 1998;
Eliev \& Shapiro 2001; Horellou \& Berge 2005;
Percival 2005; ML05).   

Also it is worth of note (Percival 2005) that
there is some confusion in the literature about
the exact form of $V_{mq}=W_{mq}$, and the
$(1+3w_q)$ term has sometimes been neglected
in the past, although it is included im more
recent work (Battye \& Weller 2004; Horellou
\& Berge 2005; Percival 2005; ML05). Owing to
a different choice of the potential, the
current expression of $V_{mq}$ coincides
with its counterpart calculated in ML05,
while the corresponding result appearing
elsewhere (e.g., Battye \& Weller 2004; 
Horellou \& Berge 2005; Percival 2005) is
different by a factor, $-1/2$.

\section{The virial theorem for matter and
quintessence two-component systems}
\label{virte}

It is worth recalling that the potential
induced by the quintessence has been 
interpreted (Sect.\,\ref{pote}) in terms
of an interaction of strenght, $(1+w_q)G$,
which depends on the inverse square distance.
Accordingly, the virial theorem for the
whole system reads (e.g., Landau \& Lifchitz
1966, Chap.\,II, \S\,10):
\begin{equation}
\label{eq:virt}
2T+\Omega=0~~;
\end{equation}
where $\Omega$ is the potential energy (e.g.,
MacMillan 1930, Chap.\,III, \S\,76; Caimmi \&
Secco 1992):
\begin{equation}
\label{eq:Ot}
\Omega=\Omega_m+W_{mq}+W_{qm}+\Omega_q~~;
\end{equation}
and $T$ is the kinetic energy.
 
On the other hand, the virial theorem 
for subsystems reads (e.g., Limber 1959;
Brosche et al. 1983; Caimmi et al. 1984;
Caimmi \& Secco 1992; Caimmi 2003):
\begin{equation}
\label{eq:viru}
2T_u+\Omega_u+V_{uv}=0~~;\qquad u=m,q~~;\qquad
v=q,m~~;
\end{equation}
where $T_u$ is the kinetic energy of $u$-th
subsystem.

At this stage, let us assume that the 
quintessence subsystem can retain some
form of kinetic energy, in such a way
it is allowed to virialize by itself,
within the tidal potential induced by
the matter subsystem.   In this view,
the substitution of Eq.\,(\ref{eq:Ot})
into (\ref{eq:virt}) yields:
\begin{equation}
\label{eq:viOW}
2T_m+2T_q+\Omega_m+\Omega_q+W_{mq}+W_{qm}=0~~;
\end{equation}
and the summation of the left-side
member of Eq.\,(\ref{eq:viru}) with
its counterpart where the indices,
$u$ and $v$, are interchanged, produces:
\begin{equation}
\label{eq:viOV}
2T_m+2T_q+\Omega_m+\Omega_q+V_{mq}+V_{qm}=0~~;
\end{equation}
finally, the combination of Eqs.\,(\ref
{eq:viOW}) and (\ref{eq:viOV}) yields:
\begin{equation}
\label{eq:VW}
V_{mq}+V_{qm}=W_{mq}+W_{qm}~~;
\end{equation}
or, without loss of generality:
\begin{lefteqnarray}
\label{eq:VWQ}
&& V_{uv}=W_{uv}+Q_{uv}~~;\qquad u=m,q
~~;\qquad v=q,m~~; \\
\label{eq:Qant}
&& Q_{mq}=-Q_{qm}~~;
\end{lefteqnarray}
where $Q_{uv}=V_{uv}-W_{uv}$ is the residual
potential energy (e.g., Caimmi \& Secco 1992).

In the special case of concentric,
spherical-symmetric, homogeneous subsystems,
the combination of Eqs.\,(\ref{eq:soii}), 
(\ref{eq:soti}), and (\ref{eq:VWQ}) yields:
\begin{equation}
\label{eq:Qij}
Q_{ij}=-\frac35(1+3w_j)\frac{GM_i^2}{R_i}
\frac m{y^3}\left(\frac54-\frac54y^2\right)~~;
\end{equation}
and the combination of Eqs.\,(\ref{eq:VWQ})
and (\ref{eq:Qant}) produces:
\begin{equation}
\label{eq:Vji}
V_{ji}=W_{ji}+Q_{ji}=W_{ji}-Q_{ij}~~;
\end{equation}
finally, using Eqs.\,(\ref{eq:soii})
and (\ref{eq:Qij}) makes Eq.\,(\ref{eq:Vji})
coincide with (\ref{eq:sotjA}).
The further restriction that the two
subsystems are bounded by a single sphere,
$R_i=R_j=R$ or $y=1$, implies $Q_{uv}=0$
and then $V_{uv}=W_{uv}$.

The virial theorem, as expressed by 
Eqs.\,(\ref{eq:virt}) and (\ref{eq:viru}),
is different from $2T+\Omega_m+2V_{mq}=0$
currently used in the literature (e.g.,
Horellou \& Berge 2005; Percival 2005),
due to the subtraction of $\Omega_q$ or
the addition of $V_{mq}$, respectively.
On the other hand, the
formulation $2T-R\partial(\Omega_m+V_
{mq}+V_{qm}+\Omega_q)/\partial R=0$
used in ML05 is valid for the whole
system, but cannot be used for
subsystems.   For further details,
see Appendix A.

In what follows, it shall be intended
that the two fluids are made of matter
$(w_m=0)$ and quintessence $(w_q=w={\rm
const})$, and fill the same volume $(R_m
=R_q=R,\,y=1)$.

\section{Fully clustered quintessence}
\label{cluq}

In the case of fully clustered quintessence,
the quintessence field responds to the infall
in the same way as matter, and the related
continuity equation reads (ML05):
\begin{equation}
\label{eq:cor}
\dot{\rho}_q+3(1+w)\frac{\dot{r}}r\rho_q=0~~;
\end{equation}
where $r$ is the radial coordinate.   An
integration from turnaround $(r=R_{\rm max})$
to a generic configuration $r=R$, using
Eq.\,(\ref{eq:Mq}) yields:
\begin{equation}
\label{eq:Mqr}
M_q(R)=M_q(R_{\rm max})\left(\frac R{R_{\rm max}}
\right)^{-3w}~~;
\end{equation}
which shows that the quintessence mass is
decreasing with radius, and \linebreak
$\lim_{R\to0}M_q(R)=0$, in the case under
consideration, $-1\le w<-1/3$.

Owing to Eqs.\,(\ref{eq:sos}), (\ref{eq:Wmq}),
and (\ref{eq:Ot}), the potential energy of the
matter subsystem within the tidal field induced
by the quintessence subststem, is:
\begin{leftsubeqnarray}
\slabel{eq:OMa}
&& \Omega_m+W_{mq}=-\frac35\frac{GM_m^2}R\left[
1+(1+3w)m_{qm}\right]~~; \\
\slabel{eq:OMb}
&& m_{qm}=\frac{M_q}{M_m}~~;
\label{seq:OM}
\end{leftsubeqnarray}
where the mass ratio, $m_{qm}$, changes with
time, according to Eq.\,(\ref{eq:Mqr}).   At
turnaround, $R=R_{\rm max}$, the assumption
of homogeneity makes total energy coincide
with potential energy i.e. null kinetic
energy (e.g., Horellou \& Berge 2005;
Percival 2005; ML05).

Owing to Eqs.\,(\ref{eq:sos}), (\ref{eq:Wmq}),
(\ref{eq:Vmq}), (\ref{eq:viru}), and (\ref
{eq:Mqr}), the potential energy of the matter
subsystem within the tidal field induced by
the quintessence subsystem at virialization,
is:
\begin{leftsubeqnarray}
\slabel{eq:virma}
&& (\Omega_m+V_{mq})_{\rm vir}=-\frac35\frac
{GM_m^2}{R_{\rm max}}\left[\frac1\eta+\frac
{(1+3w)m}{\eta^{1+3w}}\right]~~; \\
\slabel{eq:virmb}
&& \eta=\frac{R_{\rm vir}}{R_{\rm max}}~~;
\qquad m=(m_{qm})_{\rm vir}=\frac{M_q(R_{\rm 
max})}{M_m}~~;
\label{seq:virm}
\end{leftsubeqnarray}
where the indices, max and vir, denote
turnaround and virialization, respectively.

With regard to the quintessence subsystem,
the counterpart of Eqs.\,(\ref{eq:OMa}),
and (\ref{eq:virma}), via (\ref{eq:sos}),
(\ref{eq:Wmq}), (\ref{eq:Vmq}), (\ref
{eq:Ot}), (\ref{eq:OMb}), and (\ref{eq:virmb})
read:
\begin{lefteqnarray}
\label{eq:Oq}
&& \Omega_q+W_{qm}=-\frac35\frac{GM_m^2}R\left[
(1+3w)m_{qm}^2+m_{qm}\right]~~; \\
\label{eq:virq}
&& (\Omega_q+V_{qm})_{\rm vir}=-\frac35\frac
{GM_m^2}{R_{\rm max}}\left[\frac{(1+3w)m^2}
{\eta^{1+6w}}+\frac m{\eta^{1+3w}}\right]~~;
\end{lefteqnarray}
where the assumption of homogeneity makes
total energy coincide with potential energy
at turnaround.

Using Eqs.\,(\ref{eq:Ot}), (\ref{seq:OM}),
and (\ref{eq:Oq}), the potential energy of
the system at turnaround is: 
\begin{equation}
\label{eq:OMax}
\Omega_{\rm max}=-\frac35\frac{GM_m^2}{R_{\rm max}}
[1+(2+3w)m+(1+3w)m^2]~~;
\end{equation}
according to ML05.

Using Eqs.\,(\ref{eq:Wmq}), (\ref{eq:Vmq}),
(\ref{eq:Ot}), (\ref{seq:virm}), and (\ref
{eq:virq}), the potential energy of
the system at virialization is: 
\begin{equation}
\label{eq:Ovir}
\Omega_{\rm vir}=-\frac35\frac{GM_m^2}{R_{\rm max}}
\left[\frac1\eta+\frac{(2+3w)m}{\eta^{1+3w}}+\frac
{(1+3w)m^2}{\eta^{1+6w}}\right]~~;
\end{equation}
which is equivalent to:
\begin{equation}
\label{eq:Ovir2}
\Omega_{\rm vir}=-\frac35\frac{GM_m^2}{R_{\rm max}}\frac
1\eta(1+m\eta^{-3w})\left[1+(1+3w)m\eta^{-3w}\right]~~;
\end{equation}
in terms of radius and quintessence mass at
turnaround.

Keeping in mind that, in the case under discussion,
the total energy equals the potential energy at
turnaround and one half the potential energy at
virialization, the requirement of energy conservation
(e.g., Wang \& Steinhardt 1998; Weinberg \&
Kamionkonski 2003; Battye \& Weller 2003; Horellou \&
Berge 2005; ML05; but see also Percival 2005) after
combination of Eqs.\,(\ref{eq:OMax}) and (\ref
{eq:Ovir2}) yields:
\begin{equation}
\label{eq:eta}
\eta=\frac12~\frac{1+m\eta^{-3w}}{1+m}~\frac{1+(1+3w)
m\eta^{-3w}}{1+(1+3w)m}~~;
\end{equation}
where the fractional radius, $\eta=R_{\rm vir}/
R_{\rm max}$, appears on both the left-hand
and the right-hand side.   The solution of
the transcendental Eq.\,(\ref{eq:eta}) allows
the knowledge of the virialized configuration.

The counterpart of Eq.\,(\ref{eq:eta}) in
ML05, Eq.\,(16) therein, is different for
the following reason.   The expression,
$R\,\partial\Omega_{\rm vir}/\partial R$,
has been calculated at constant quintessence
mass for an assigned configuration in the
current approach, while the quintessence
mass, $M_q$, has been derived with respect
to the radius, $R$, in ML05.

The fractional radius, $\eta=R_{\rm vir}/ 
R_{\rm max}$, is necessarily restricted
within the range, $0\le\eta\le1$.   In fact,
distances cannot be negative and, on the
other hand, $\eta>1$ would contradict the
definition of turnaround radius, $R_{\rm
max}$.   To gain more insight, let us write
Eq.\,(\ref{eq:eta}) under the form:
\begin{leftsubeqnarray}
\slabel{eq:fxa}
&& \eta=\frac12f(x)=\frac12~\frac{1+x}{1+m}~
\frac{1+(1+3w)x}{1+(1+3w)m}~~; \\
\slabel{eq:fxb}
&& x=m\eta^{-3w}~~;\qquad0\le x\le m~~;\qquad0\le
f(x)\le2~~;
\label{seq:fx}
\end{leftsubeqnarray}
where the function, $f(x)$, is studied in
Appendix B.   In the limiting case of a
vanishing quintessence field, $m\to0$,
$x\to0$, and Eq.\,(\ref{eq:fxa}) reduces
to $\eta=1/2$, which is the known result
in absence of dark energy.

The combination of Eqs.\,(\ref{eq:fxa})
and (\ref{eq:fxb}) yields:
\begin{equation}
\label{eq:Fx}
\left(\frac xm\right)^{-1/(3w)}=\frac12~\frac
{1+x}{1+m}~\frac{1+(1+3w)x}{1+(1+3w)m}~~;
\end{equation}
and the virialized configuration is 
determined by the intersection point
between the curve on the left-hand and
right-hand side, respectively.

To gain further insight, let us take
into consideration a few special cases,
namely $w=-1$, $-$2/3, $-$1/2, $-$1/3,
the last to be thought of as an interesting
limiting situation.   The general case,
$-1\le w<-1/3$, is expected to show similar
properties as in the closest among the above
mentioned particular situations.

\begin{figure}
\centerline{\psfig{file=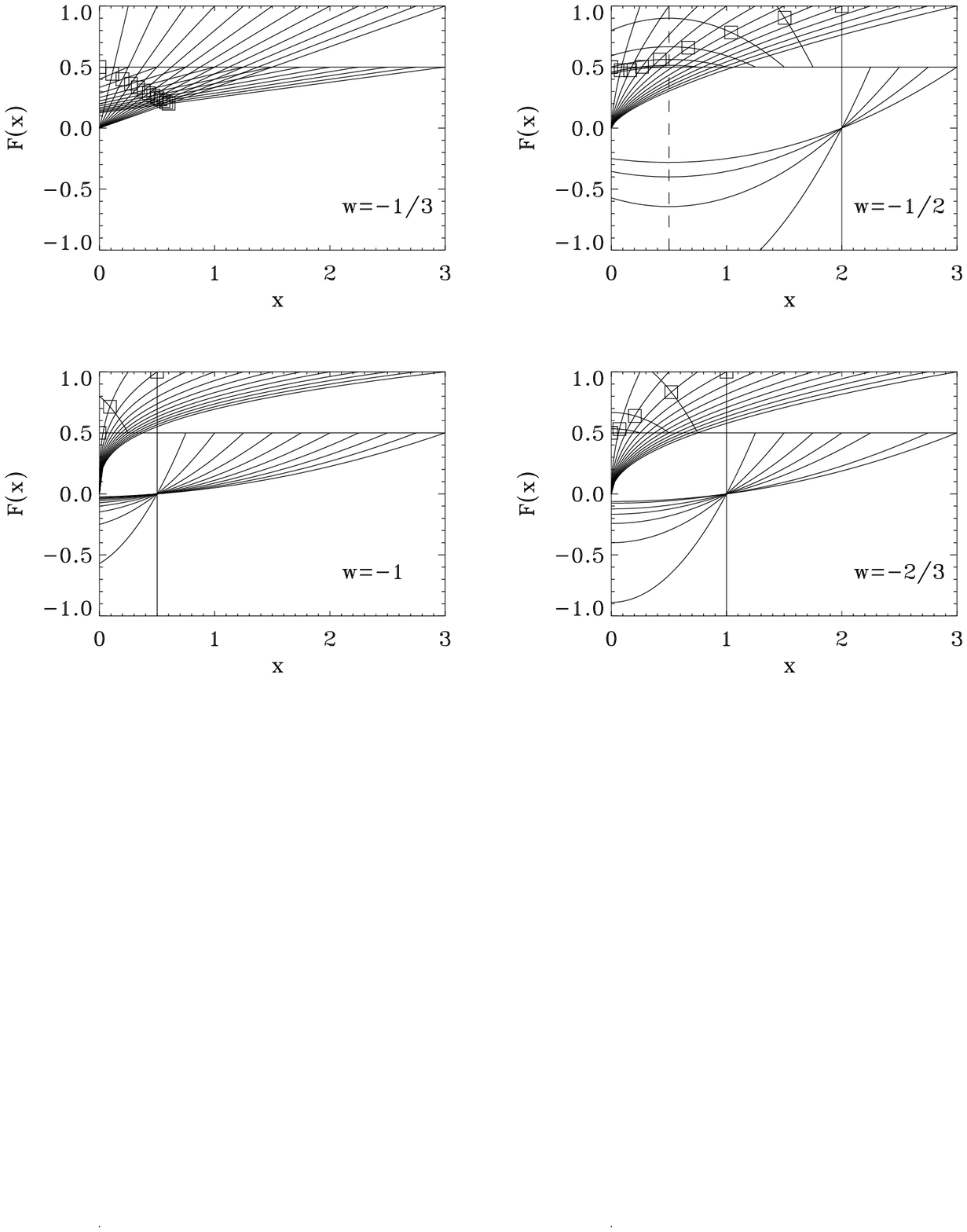,height=100mm,width=90mm}}
\caption{The fractional radius, $\eta=R_{\rm vir}/
R_{\rm max}$, for different values of the fractional
mass, $m=M_q(R_{\rm max})/M_m$, with regard to global
energy conservation and special choices of the 
equation of state parameter, $w$, from top left in the
clockwise sense: $-$1/3, $-$1/2, $-$2/3, $-$1.
The value of $m$ related to each case can be
read as the abscissa of the ending point of full
curves, which occurs at $F(x)=1/2$.   The intersection
of the two curves on the left-hand and right-hand side
of Eq.\,(\ref{eq:Fx}), related to a virialized
configuration, is marked by an open square.
Curves corresponding to $F(x)=(x/m)^{-1/(3w)}$ start
from the origin.   Curves corresponding to $F(x)=(1/2)
f(x)$ end at $F(x)=1/2$, according to Eq.\,(\ref
{eq:Fx}).   The horisontal axis corresponds to
dominant quintessence mass, $m\to+\infty$.}
\label{f:etag}    
\end{figure}

In the special case, $w=-1/3$, Eq.\,(\ref
{eq:Fx}) reduces to:
\begin{equation}
\label{eq:F1}
\frac xm=\frac12\frac{1+x}{1+m}~~;
\end{equation}
the solution of which is:
\begin{equation}
\label{eq:eta1}
\eta=\frac xm=\frac1{m+2}~~;
\end{equation}
via Eq.\,(\ref{eq:fxb}).   The fractional
radius, $\eta$, for different values of
the fractional mass, $m$, is represented
in Fig.\,\ref{f:etag}, top left.

In the special case, $w=-1/2$, Eq.\,(\ref
{eq:Fx}) reduces to:
\begin{equation}
\label{eq:F2}
\left(\frac xm\right)^{2/3}=\frac12\frac{2+x-x^2}
{2+m-m^2}~~;
\end{equation}
the solution of which is determined by the
intersection between curves on the left-hand
and right-hand side.   The fractional
radius, $\eta$, for different values of
the fractional mass, $m$, is represented
in Fig.\,\ref{f:etag}, top right, where
intersections are marked by open squares.

In the special case, $w=-2/3$, Eq.\,(\ref
{eq:Fx}) reduces to:
\begin{equation}
\label{eq:F3}
\left(\frac xm\right)^{1/2}=\frac12\frac{1-x^2}
{1-m^2}~~;
\end{equation}
the solution of which is determined by the
intersection between curves on the left-hand
and right-hand side.   The fractional
radius, $\eta$, for different values of
the fractional mass, $m$, is represented
in Fig.\,\ref{f:etag}, bottom right, where
intersections are marked by open squares.

In the special case, $w=-1$, Eq.\,(\ref
{eq:Fx}) reduces to:
\begin{equation}
\label{eq:F4}
\left(\frac xm\right)^{1/3}=\frac12\frac{1-x-2x^2}
{1-m-2m^2}~~;
\end{equation}
the solution of which is determined by the
intersection between curves on the left-hand
and right-hand side.   The fractional
radius, $\eta$, for different values of
the fractional mass, $m$, is represented
in Fig.\,\ref{f:etag}, bottom left, where
intersections are marked by open squares.

An inspection of Fig.\,\ref{f:etag} shows
the following features.
\begin{description}
\item[\rm{(i)}\hspace{3.5mm}] For assigned
turnaround radius, the virialization radius
is a decreasing function of the quintessence
mass for $w=-1/3$, and an increasing function
for $w\le-2/3$.   The trend is non monotonic
for $w=-1/2$, with the occurrence of a minimum
point.
\item[\rm{(ii)}~~] For assigned equation of
state parameter, $w$, a virialized configuration
is allowed for turnaround fractional mass within
the range, $0\le m\le m_0$, where $m_0$ is the
value for which the function on the right-hand
side of Eq.\,(\ref{eq:Fx}) is divergent.   In
the limit, $m\to m_0$, density perturbations
turn around and virialize at infinite radius,
while density perturbations where $m>m_0$ cannot
virialize.
\item[\rm{(iii)}~] The threshold fractional mass
ranges from $m_0\to+\infty$ $(w=-1/3)$ to $m_0=
2$ $(w=-1/2)$; $m_0=1$ $(w=-2/3)$; and $m_0=1/2$
$(w=-1)$.   Accordingly, the minimum amount of
quintessence within a density perturbation,
necessary to prevent matter virialization, is
an increasing function of the equation of state
parameter, $w$.
\end{description}

The above results rely on the assumption, that
both energy conservation and virialization hold
for the whole system (ML05).   If, on the other
hand, only the matter virializes i.e. the kinetic
energy of the quintessence subsystem is null, the
procedure should be repeated using Eqs.\,(\ref
{seq:OM}) and (\ref{seq:virm}) instead of (\ref
{eq:OMax}) and (\ref{eq:Ovir}).   The result is:
\begin{leftsubeqnarray}
\slabel{eq:eta1Ca}
&& \eta=\left(\frac xm\right)^{-1/(3w)}=\frac12
\frac{1+(1+3w)x}{1+(1+3w)m}~~; \\
\slabel{eq:eta1Cb}
&& x=m\eta^{-3w}~~;\qquad0\le x\le m~~;
\label{seq:eta1C}
\end{leftsubeqnarray}
where the virialized configuration is
defined by the intersection of the
curves on the left-hand and right-hand
side of Eq.\,(\ref{eq:eta1Ca}), the
latter being straight lines.

The counterpart of Eq.\,(\ref{eq:eta1Ca})
in ML05, Eq.\,(17) therein, is different
for the following reason.   Though no
indication is provided in ML05 on how
Eq.\,(17) therein has been derived, it
can be seen that it is sufficient to
omit the terms $U_{12}$ and $U_{22}$
($V_{mq}$ and $\Omega_q$ in the current
notation) in the expression of the
potential energy, Eq.\,(4) therein.
On the contrary, using Eqs.\,(\ref
{seq:OM}) and (\ref{seq:virm}) implies
the omission of $V_{qm}$ and $\Omega_q$,
which explains the different results.

In the special case, $w=-1/3$, Eq.\,(\ref
{eq:eta1Ca}) reduces to $\eta=1/2$, which
is the result for matter universes.

\begin{figure}
\centerline{\psfig{file=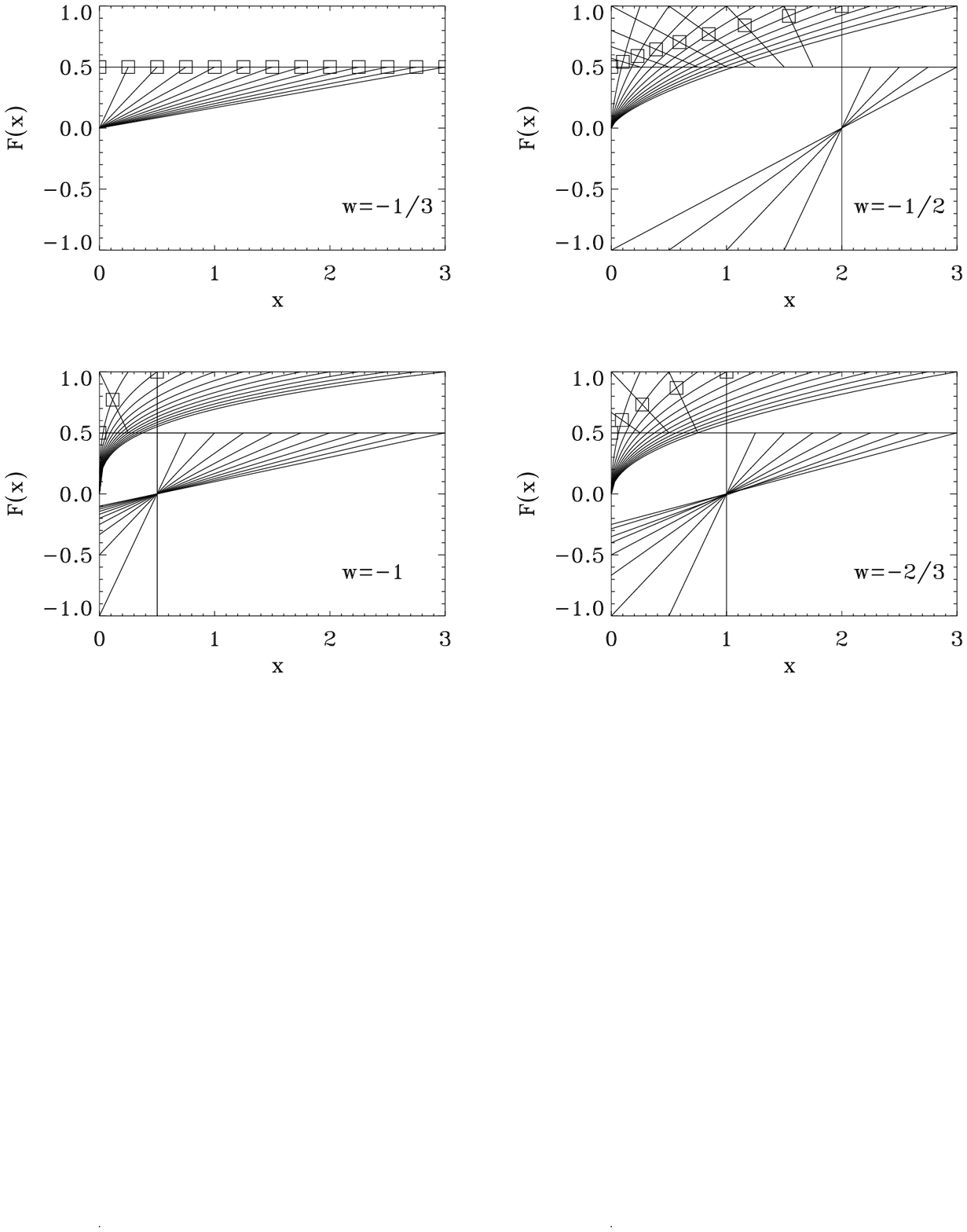,height=100mm,width=90mm}}
\caption{The fractional radius, $\eta=R_{\rm vir}/
R_{\rm max}$, for different values of the fractional
mass, $m=M_q(R_{\rm max})/M_m$, with regard to partial
energy conservation and special choices of the 
equation of state parameter, $w$, from top left in the
clockwise sense: $-$1/3, $-$1/2, $-$2/3, $-$1.
The value of $m$ related to each case can be
read as the abscissa of the ending point of full
curves, which occurs at $F(x)=1/2$.   The intersection
of the two curves on the left-hand and right-hand side
of Eq.\,(\ref{eq:eta1Ca}), related to a virialized
configuration, is marked by an open square.
Curves corresponding to $F(x)=(x/m)^{-1/(3w)}$ start
from the origin.   Straight lines corresponding to
$F(x)=(1/2)[1+(1+3w)x]/[1+(1+3w)m]$ end at $F(x)=
1/2$, according to Eq.\,(\ref{eq:eta1Ca}).   The
horisontal axis corresponds to
dominant quintessence mass, $m\to+\infty$.}
\label{f:etap}    
\end{figure}

In the special case, $w=-1/2$, Eq.\,(\ref
{eq:eta1Ca}) reduces to:
\begin{equation}
\label{eq:F1C}
\left(\frac xm\right)^{2/3}=\frac12\frac{2-x}
{2-m}~~;
\end{equation}
the solution of which is determined by the
intersection between curves on the left-hand
and right-hand side.   The fractional
radius, $\eta$, for different values of
the fractional mass, $m$, is represented
in Fig.\,\ref{f:etap}, top right, where
intersections are marked by open squares.

In the special case, $w=-2/3$, Eq.\,(\ref
{eq:eta1Ca}) reduces to:
\begin{equation}
\label{eq:F2C}
\left(\frac xm\right)^{1/2}=\frac12\frac{1-x}
{1-m}~~;
\end{equation}
the solution of which is determined by the
intersection between curves on the left-hand
and right-hand side.   The fractional
radius, $\eta$, for different values of
the fractional mass, $m$, is represented
in Fig.\,\ref{f:etap}, bottom right, where
intersections are marked by open squares.

In the special case, $w=-1$, Eq.\,(\ref
{eq:eta1Ca}) reduces to:
\begin{equation}
\label{eq:F3C}
\left(\frac xm\right)^{1/3}=\frac12\frac{1-2x}
{1-2m}~~;
\end{equation}
the solution of which is determined by the
intersection between curves on the left-hand
and right-hand side.   The fractional
radius, $\eta$, for different values of
the fractional mass, $m$, is represented
in Fig.\,\ref{f:etap}, bottom left, where
intersections are marked by open squares.

An inspection of Fig.\,\ref{f:etap} shows
similar features as in Fig.\,\ref{f:etag}
(see above), with the following difference.
For an assigned turnaround radius, the
virialization radius is an increasing
function of the quintessence mass for
$w<-1/3$, and independent of the
quintessence mass (as in matter universes)
for $w=-1/3$.   Accordingly, a monotonic
trend is exhibited.

The results plotted in Fig.\,\ref{f:etag}
and Fig.\,\ref{f:etap} imply the assumption of
energy conservation with regard to the whole
system and to the matter subsystem, respectively,
the latter requirement being more restrictive
than the former.   For a more detailed discussion
on energy conservation in density perturbations
with matter and dark energy, see ML05 and Percival
(2005).

The comparison of the current results 
(Figs.\,\ref{f:etag}-\ref{f:etap}) with their
counterparts in ML05 (Fig.\,2 therein, related
to $w=-4/5$) shows that the trend of the
fractional radius, $\eta$, as a function of
the fractional mass, $m$, is increasing in
both cases with regard to the whole system,
while is increasing in the current paper and
decreasing in ML05 with regard to the matter
subsystem.   A closer agreement could occur
within the range, $-1/2\le w<-1/3$.   The
reasons of the above mentioned discrepancy
are due to different assumptions made in
the current paper and in ML05, as explained
above.

\section{Partially clustered quintessence}
\label{uncq}

In the case of partially clustered quintessence,
the quintessence field responds to the infall
to a lesser extent with respect to the matter.
In the limiting situation of unclustered
quintessence, the related
continuity equation reads (ML05):
\begin{equation}
\label{eq:coa}
\dot{\rho}_q+3(1+w)\frac{\dot{a}}a\rho_q=0~~;
\end{equation}
where $a$ is the cosmological scale factor,
which is related to the redshift, $z$, via
$a\propto(1+z)^{-1}$.
In the special case, $w=-1$, Eqs.\,(\ref{eq:cor})
and (\ref{eq:coa}) coincide, yielding $\rho_q=
{\rm const}$ or, in other terms, a cosmological
constant.

An integration from turnaround $(r=R_{\rm max})$
to a generic configuration $r=R$, using
Eq.\,(\ref{eq:Mq}) yields:
\begin{leftsubeqnarray}
\slabel{eq:Mqaa}
&& M_q(R)=M_q(R_{\rm max})\left(\frac R{R_{\rm max}}
\right)^3\left(\frac a{a_{\rm max}}\right)^
{-3(1+w)}~~; \\
\slabel{eq:Mqab}
&& \frac a{a_{\rm max}}=\frac{1+z_{\rm max}}
{1+z}~~;
\label{seq:Mqa}
\end{leftsubeqnarray}
which shows that the quintessence mass is
decreasing with radius, and \linebreak
$\lim_{R\to0}M_q(R)=0$, in the case under
consideration, $-1\le w<-1/3$.

In the general case of partially clustered
quintessence, it may safely be expected that
the virialized configuration of matter
subsystem lies between the limiting situation
of fully clustered and unclustered quintessence,
respectively.   It is worth remembering that
energy conservation can no longer be assumed,
as in presence of unclustered quintessence
(e.g., Horellou \& Berge 2005; Percival 2005;
ML05).   Without loss of generality, let us
suppose that a specified virialized configuration
is attained along the following steps: (i) from
turnaround to virialization related to fully
clustered quintessence, and (ii) from
virialization related to fully clustered 
quintessence to virialization related to
partially clustered quintessence.

To perform the transition between the above
mentioned steps, the quintessence mass must
instantaneously change from $M_q(R_{\rm vir}
^\prime)$ to $M_q(R_{\rm vir})$, where the presence
and the absence of the prime denotes the virialized
configuration related to fully or partially
clustered quintessence, respectively.

The changes in self, tidal, and interaction
potential energy, keeping in mind Eqs.\,(\ref
{eq:sos}), (\ref{eq:Wmq}), and (\ref{eq:Vmq}),
are:
\begin{lefteqnarray}
\label{eq:DOm1}
&& \Delta\Omega_m=-\frac35\frac{GM_m^2}{R_{\rm
vir}^\prime}+\frac35\frac{GM_m^2}{R_{\rm vir}^
\prime}~~; \\
\label{eq:DVmq1}
&& \Delta V_{mq}=\Delta W_{mq}=-\frac35(1+3w)
\frac{GM_mM_q(R_{\rm vir}^\prime)}{R_{\rm 
vir}^\prime} \nonumber \\
&& \phantom{\Delta V_{mq}=\Delta W_{mq}=}+
\frac35(1+3w)\frac{GM_mM_q(R_{\rm vir})}{R_{\rm 
vir}^\prime}~~; \\
\label{eq:DVqm1}
&& \Delta V_{qm}=\Delta W_{qm}=-\frac35
\frac{GM_mM_q(R_{\rm vir}^\prime)}{R_{\rm 
vir}^\prime} \nonumber \\
&& \phantom{\Delta V_{mq}=\Delta W_{mq}=}+
\frac35\frac{GM_mM_q(R_{\rm vir})}{R_{\rm 
vir}^\prime}~~; \\
\label{eq:DOq1}
&& \Delta\Omega_q=-\frac35(1+3w)\frac{GM_q
^2(R_{\rm vir}^\prime)}{R_{\rm vir}^\prime}+
\frac35(1+3w)\frac{GM_q^2(R_{\rm vir})}{R_
{\rm vir}^\prime}~~;
\end{lefteqnarray}
from which the changes in total energy within
the matter subsystem and the whole system,
can be deduced.

To this aim, let us define the fractional masses
and the fractional radii:
\begin{lefteqnarray}
\label{eq:Dmu}
&& \mu=\frac{M_q(R_{\rm vir})}{M_q(R_{\rm max})}~~;
\qquad\Delta\mu=\frac{M_q(R_{\rm vir}^\prime)-M_q
(R_{\rm vir})}{M_q(R_{\rm max})}~~; \\
\label{eq:Deta}
&& \eta=\frac{R_{\rm vir}}{R_{\rm max}}~~;\qquad
\Delta\eta=\frac{R_{\rm vir}^\prime-R_{\rm vir}}
{R_{\rm max}}~~;
\end{lefteqnarray}
and the combination of Eqs.\,(\ref{eq:Mqr}), 
(\ref{eq:Dmu}), and (\ref{eq:Deta}) yields:
\begin{lefteqnarray}
\label{eq:muD}
&& \mu+\Delta\mu=\frac{M_q(R_{\rm vir}^\prime)}
{M_q(R_{\rm max})}=\left(\frac{R_{\rm vir}^\prime}
{R_{\rm max}}\right)^{-3w}=\eta_{\rm FC}^{-3w}~~; \\
\label{eq:etaD}
&& \eta+\Delta\eta=\frac{R_{\rm vir}^
\prime}{R_{\rm max}}=\eta_{\rm FC}~~;
\end{lefteqnarray}
in the special case of fully clustered
quintessence, $\Delta\mu=0$, the results of
Sect.\,\ref{cluq} continue to hold provided
$\eta$ is replaced by $\eta_{\rm FC}$ therein.

The substitution of Eqs.\,(\ref{seq:virm}) and
(\ref{eq:Dmu})-(\ref{eq:etaD}) into (\ref
{eq:DOm1})-(\ref{eq:DOq1}) produces:
\begin{lefteqnarray}
\label{eq:DOm2}
&& \Delta\Omega_m=0~~; \\
\label{eq:DVmq2}
&& \Delta V_{mq}=\Delta W_{mq}=-\frac35(1+3w)
\frac{GM_m^2}{R_{\rm vir}^\prime}m\Delta\mu~~; \\
\label{eq:DVqm2}
&& \Delta V_{qm}=\Delta W_{qm}=-\frac35
\frac{GM_m^2}{R_{\rm vir}^\prime}m\Delta\mu~~; \\
\label{eq:DOq2}
&& \Delta\Omega_q=-\frac35(1+3w)\frac{GM_m
^2}{R_{\rm vir}^\prime}m^2[2\mu\Delta\mu+
(\Delta\mu)^2]~~;
\end{lefteqnarray}
in terms of virialized configurations related
to fully clustered quintessence and to the
change, $\Delta\mu$.

At this stage, the virialized configuration
of the matter subsystem can be determined
using a similar procedure with respect to
the special case of fully clustered quintessence.
The change in total energy, related to the 
transition from virialized configurations
where the quintessence is fully clustered
$(R=R_{\rm vir}^\prime)$ to virialized 
configurations where the quintessence is 
partially clustered $(R=R_{\rm vir})$,
reads: 
\begin{leftsubeqnarray}
\slabel{eq:DEa}
&& \frac12\Omega(R_{\rm vir}^\prime)+
\Delta\Omega=\frac12\Omega(R_{\rm vir})~~; \\
\slabel{eq:DEb}
&& \Delta\Omega=\Delta\Omega_m+\Delta V_{mq}+
\Delta V_{qm}+\Delta\Omega_q~~;
\label{seq:DE}
\end{leftsubeqnarray}
and the combination of Eqs.\,(\ref{eq:Ovir}),
(\ref{eq:Dmu})-(\ref{eq:etaD}), and (\ref
{eq:DOm2})-(\ref{seq:DE}) yields:
\begin{lefteqnarray}
\label{eq:Ovp}
&& \Omega(R_{\rm vir}^\prime)=-\frac35\frac
{GM_m^2}{R_{\rm max}}\left[\frac1{\eta_{\rm 
FC}}+\frac{(2+3w)m}{\eta_{\rm FC}^{1+3w}}+
\frac{(1+3w)m^2}{\eta_{\rm FC}^{1+6w}}\right]~~; \\
\label{eq:Ov}
&& \Omega(R_{\rm vir})=-\frac35\frac
{GM_m^2}{R_{\rm max}}\left[\frac1\eta 
+\frac{(2+3w)m\mu}\eta+
\frac{(1+3w)m^2\mu^2}\eta\right]~~; \\
\label{eq:DOv}
&& \Delta\Omega=-\frac35\frac
{GM_m^2}{R_{\rm max}}\left[(2+3w)m+
(1+3w)m^2(2\mu+\Delta\mu)\right]\frac
{\Delta\mu}{\eta_{\rm FC}}~~;
\end{lefteqnarray}
in terms of radius and quintessence mass
at turnaround.

Keeping in mind that, in the special case
of fully clustered quintessence, the total
energy equalizes the potential energy at
turnaround and one half the potential energy
at virialization, the substitution of
Eqs.\,(\ref{eq:OMax}), (\ref{eq:Ov}), and
(\ref{eq:DOv}) into (\ref{eq:DEa}) produces:
\begin{lefteqnarray}
\label{eq:etp1}
&& [1+(2+3w)m+(1+3w)m^2]+[(2+3w)m+(1+3w)m^2
(2\eta_{\rm FC}^{-3w}-\Delta\mu)]\frac{\Delta
\mu}{\eta_{\rm FC}} \nonumber \\
&& =\frac12[1+(2+3w)m\mu+(1+3w)m^2\mu^2]\frac1
\eta~~;
\end{lefteqnarray}
where, using Eq.\,(\ref{eq:muD}) and performing
some algebra, the term within brackets on the 
right-hand side of Eq.\,(\ref{eq:etp1}) may be 
cast into the form:
\begin{leftsubeqnarray}
\slabel{eq:etp2a}
&& 1+(2+3w)m\mu+(1+3w)m^2\mu^2= \nonumber \\
&& \qquad1+(2+3w)m\eta_{\rm
FC}^{-3w}+(1+3w)m^2\eta_{\rm FC}^{-6w}-\phi(w,m,
\eta_{\rm FC},\Delta\mu)~~; \\
\slabel{eq:etp2b}
&& \phi(w,m,\eta_{\rm FC},\Delta\mu)= \nonumber \\
&& \qquad m[(2+3w)+2(1+3w)m\eta_{\rm FC}^{-3w}-
(1+3w)m\Delta\mu]\Delta\mu~~;
\label{seq:etp2}
\end{leftsubeqnarray}
and the combination of Eqs.\,(\ref{eq:etp1})
and (\ref{seq:etp2}) yields:
\begin{equation}
\label{eq:etp3}
\eta=\frac12\frac{[1+(2+3w)m\eta_{\rm FC}^
{-3w}+(1+3w)m^2\eta_{\rm FC}^{-6w}]-\phi}
{[1+(2+3w)m+(1+3w)m^2]+\eta_{\rm FC}^{-1}
\phi}~~;
\end{equation}
which depends on the parameters, $w$, $m$,
$\eta_{\rm FC}$, and $\Delta\mu$.

In the special case of fully clustered
quintessence, $\Delta\mu=0$ i.e. $\mu=
\eta_{\rm FC}^{-3w}$, and Eq.\,(\ref
{seq:etp2}) reduces to (\ref{eq:eta})
where $\eta$ has to be replaced by
$\eta_{\rm FC}$.   The solution of
the transcendental Eq.\,(\ref{eq:etp3})
allows the knowledge of the virialized
configuration for fixed $\Delta\mu$.

The counterpart of Eq.\,(\ref{eq:etp3})
in ML05, Eq.\,(23) therein, is different
for reasons discussed in Sect.\,\ref{cluq}
and, in addition, due to a different
choice of the parameter related to partial
clustering, $\gamma$ therein instead of
$\Delta\mu$ or $\Delta\eta$, according to
Eq.\,(\ref{eq:muD}).

The fractional radius, $\eta=R_{\rm vir}/
R_{\rm max}$, is necessarily restricted
within the range, $0\le\eta\le1$.   In
fact, distances cannot be negative and,
on the other hand, $\eta>1$ would contradict
the definition of turnaround radius,
$R_{\rm max}$.   In addition, the
particularization of Eqs.\,(\ref{eq:Mqr})
and (\ref{seq:Mqa}) to the related virialized
configurations,  shows that $M_q(R_{\rm vir})
\le M_q(R_{\rm vir}^\prime)$ within the range
of interest, $-1\le w<-1/3$, which implies
$\Delta\mu\ge0$ via Eq.\,(\ref{eq:Dmu}).
Accordingly, $0\le\Delta\mu\le\eta_{\rm FC}^
{-3w}$, and the sign of $\phi$ is opposite
to the sign of $\Delta\Omega$ via Eq.\,(\ref
{eq:DOv}) which, in turn, is equal to the
sign of $\eta-\eta_{\rm FC}$.

To gain more insight, let us express the
increment, $\Delta\mu$, in terms of the
variable, $\eta_{\rm FC}$, and a degree
of quintessence de-clustering, $\zeta$, as:
\begin{equation}
\label{eq:zita}
\Delta\mu=\zeta\eta_{\rm FC}^{-3w}~~;\qquad
0\le\zeta\le1~~;
\end{equation}
the substitution of Eq.\,(\ref{eq:zita})
into (\ref{eq:etp2b}) yields:
\begin{equation}
\label{eq:phi}
\phi(w,m,\eta_{\rm FC},\Delta\mu)=(2+3w)m\zeta
\eta_{\rm FC}^{-3w}+(1+3w)m^2\zeta(2-\zeta)
\eta_{\rm FC}^{-6w}~~;
\end{equation}
where the effect of quintessence partial
clustering is expressed by the parameter,
$\zeta$, where the limit of fully
clustered and fully de-clustered quintessence
corresponds to $\zeta=0$ and $\zeta=1$,
respectively.   The latter, of course,
has no physical meaning, as quintessence
cannot be devoided from the volume filled
by the matter subsystem.   On the other
hand, it makes an upper limit to the
quintessence de-clustering parameter,
$\zeta$.   Further details on the function,
$\phi$, are given in Appendix C.

The above results may be summarized as
follows.   Given a density perturbation
with assigned value of quintessence
equation of state parameter, $w$, $-1
\le w<-1/3$, quintessence to matter
mass ratio at turnaround, $m$,
virialized to turnaround size ratio
in the special case of fully clustered
quintessence, $\eta_{\rm FC}$, $0\le
\eta_{\rm FC}\le1$, and degree of
quintessence de-clustering, $\zeta$,
$0\le\zeta\le1$, the related
virialized to turnaround size ratio
in the case of partially clustered
quintessence, $\eta$, is expressed by
Eqs.\,(\ref{eq:etp3}), (\ref{eq:zita}),
and (\ref{eq:phi}).

Let us repeat that the special case of
fully de-clustered quintessence, $\zeta
=1$, makes a convenient upper limit but,
on the other hand, it has little physical
meaning.   A true upper limit is related
to the special case of unclustered
quintessence, $\zeta=\zeta_{\rm unc}$,
where the combination of Eqs.\,(\ref
{seq:Mqa}), (\ref{eq:Dmu}), (\ref{eq:Deta}),
(\ref{eq:muD}), and (\ref{eq:zita}) 
yields:
\begin{equation}
\label{eq:etau}
\eta=(1-\zeta_{\rm unc})^{1/3}\eta_{\rm FC}^{-w}
\left(\frac{1+z_{\rm max}}{1+z}\right)^
{1+w}~~;
\end{equation}
which, in addition to Eq.\,(\ref{eq:etp3}),
makes a further constraint for determining
the virialized configuration, with regard
to selected turnaround and virialization
epoch.

To get further insight, let us express the
fractional radius, $\eta$, in terms of the
parameter, $x=m\eta_{\rm FC}^{-3w}$.
Accordingly, Eq.\,(\ref{eq:etp3}) reads:
\begin{equation}
\label{eq:etp4}
\eta=\frac12\frac{[1+(2+3w)x+(1+3w)x^2]-
\phi(x;w,\zeta)}
{[1+(2+3w)m+(1+3w)m^2]+(x/m)^{1/(3w)}
\phi(x;w,\zeta)}~~;
\end{equation}
where the function, $\phi(x;w,\zeta)$,
is expressed by Eq.\,(\ref{eq:phx}),
Appendix C.

\begin{figure}
\centerline{\psfig{file=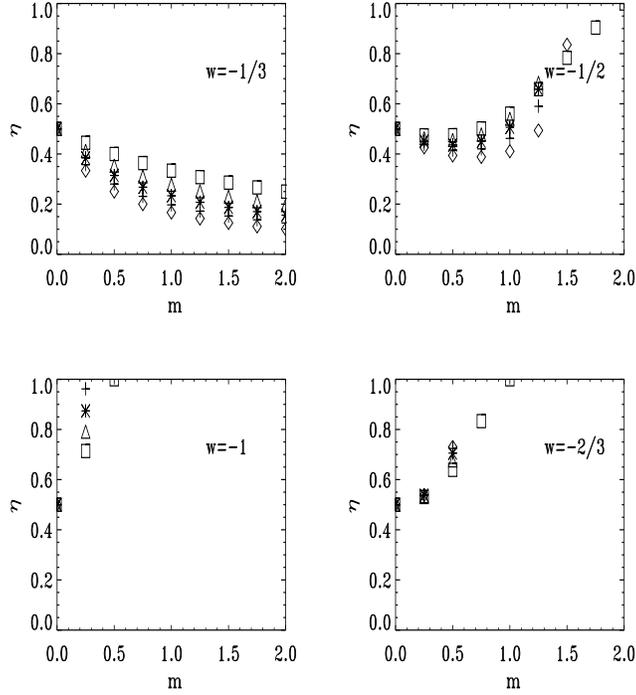,height=100mm,width=90mm}}
\caption{The fractional radius, $\eta=R_{\rm vir}/
R_{\rm max}$, with regard to global energy conservation,
for different values of the fractional
mass, $m=M_q(R_{\rm max})/M_m$, the quintessence
de-clustering parameter, $\zeta$, and the quintessence
equation of state parameter, $w$.   From top left in the
clockwise sense: $w=-$1/3, $-$1/2, $-$2/3, $-$1.
Caption of symbols: squares, $\zeta=0$ (fully clustered
quintessence, as in Fig.\,\ref{f:etag}); triangles,
$\zeta=1/4$; asterisks, $\zeta=1/2$; crosses, $\zeta=
3/4$; diamonds, $\zeta=1$ (fully de-clustered quintessence).
The value of $m$ related to each symbol, starting from
the left, is $m=i/4$, $i=0,1,2,...,$ and the special
case, $i=0$, corresponds to a vanishing quintessence.}
\label{f:etam}    
\end{figure}

The dependence, $\eta=\eta(m)$, is
represented in Fig.\,\ref{f:etam} with
regard to same cases as in Fig.\,\ref
{f:etag}, but different degrees of
quintessence de-clustering, $\zeta=0$
(squares), 0.25 (triangles), 0.5
(asterisks), 0.75 (crosses), and
1 (diamonds). 
The starting point on the left, (0,0.5), marks the
limiting situation of vanishing quintessence, and
necessarily coincides in all cases.   For sufficiently
high values of quintessence equation of state parameter,
$w\leq-1/3$, lower curves correspond to less clustered
quintessence and vice versa.   For sufficiently low
values of quintessence equation of state parameter,
$-1\le w\le-2/3$, lower curves correspond to more
clustered quintessence and vice versa.   For intermediate
values of quintessence equation of state parameter,
$w\approx-1/2$, different curves intersect one with
the other, but a minimum point occurs in the whole
range between fully clustered $(\zeta=0)$ and fully
de-clustered $(\zeta=1)$ quintessence.   In any case,
the general trend remains unchanged with respect to
the limiting situation of fully clustered quintessence.

The above results rely on the assumption, that both
energy conservation and virialization hold for the
whole system (ML05).   If, on the other hand, only
the matter virializes i.e. the kinetic energy of
the quintessence system is null, the procedure
should be repeated with regard to the matter
subsystem only.   Accordingly, Eqs.\,(\ref{seq:DE}),
(\ref{eq:Ovp}), (\ref{eq:Ov}), and (\ref{eq:DOv})
are turned into:
\begin{lefteqnarray}
\label{eq:DEm1}
&& E_m(R_{\rm vir}^\prime)+\Delta E_m=E_m(R_{\rm vir})~~; \\
\label{eq:DEm2}
&& \Delta E_m=\Delta\Omega_m+\Delta V_{mq}~~; \\
\label{eq:Ovpm}
&& 2E_m(R_{\rm vir}^\prime)=-\frac35\frac{GM_m^2}{R_{\rm max}}
\frac1{\eta_{\rm FC}}\left[1+\frac{(1+3w)m}{\eta_{\rm FC}^{3w}}
\right]~~; \\
\label{eq:Ovm}
&& 2E_m(R_{\rm vir})=-\frac35\frac{GM_m^2}{R_{\rm max}}\frac
1\eta[1+(1+3w)m\mu]~~; \\
\label{eq:DOvm}
&& \Delta E_m=-\frac35\frac{GM_m^2}{R_{\rm max}}(1+3w)
\frac{m\Delta\mu}{\eta_{\rm FC}}~~;
\end{lefteqnarray}
where $E_m=(\Omega_m+W_{mq})/2=(\Omega_m+V_{mq})/2$
is the total energy of the virialized matter
subsystem, and $\Delta E_m$ is the energy change
due to the transition from a virialized configuration
where the quintessence is fully clustered, to its
counterpart where the quintessence is partially
clustered.   Using energy conservation in the
former alternative, the substitution of Eqs.\,(\ref
{eq:Ovpm}), (\ref{eq:Ovm}), and (\ref{eq:DOvm})
into (\ref{eq:DEm1}) yields after some algebra:
\begin{lefteqnarray}
\label{eq:etaf1}
&& \eta=\frac12\frac{[1+(1+3w)m\eta_{\rm FC}^{-3w}]-
\phi}{[1+(1+3w)m]+\eta_{\rm FC}^{-1}\phi}~~; \\
\label{eq:fip1}
&& \phi(w,m,\zeta)=(1+3w)m\zeta\eta_{\rm FC}^{-3w}~~;
\end{lefteqnarray}
where Eq.\,(\ref{eq:muD}) has also been used.

In terms of the parameter, $x=m\eta_{\rm FC}^{-3w}$,
Eqs.\,(\ref{eq:etaf1}) and (\ref{eq:fip1}), translate
into:
\begin{lefteqnarray}
\label{eq:etaf2}
&& \eta=\frac12\frac{[1+(1+3w)x]-\phi}{[1+(1+3w)m]+
(x/m)^{1/(3w)}\phi}~~; \\
\label{eq:fip2}
&& \phi(x,w,\zeta)=(1+3w)\zeta x~~;
\end{lefteqnarray}
in the limit of fully clustered quintessence,
$\zeta=0$, $\phi=0$, Eq.\,(\ref{eq:etaf2})
coincides with Eq.\,(\ref{eq:eta1Ca}).

\begin{figure}
\centerline{\psfig{file=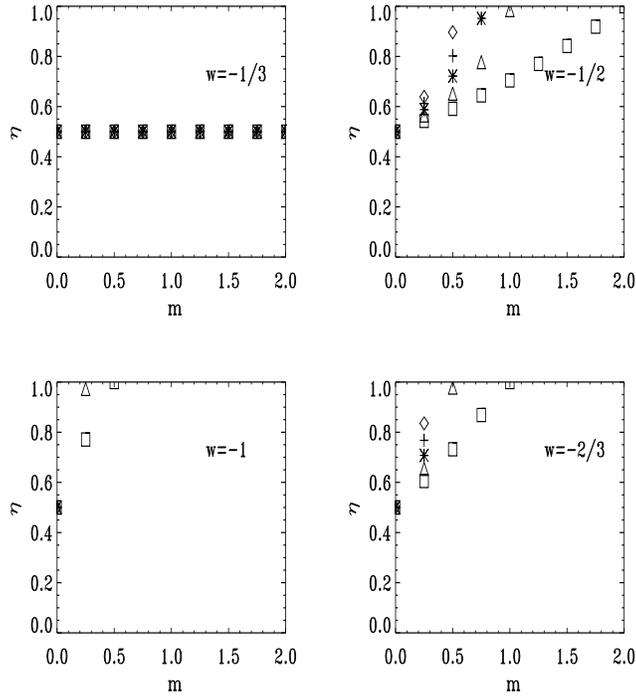,height=100mm,width=90mm}}
\caption{The fractional radius, $\eta=R_{\rm vir}/
R_{\rm max}$, for different values of the fractional
mass, $m=M_q(R_{\rm max})/M_m$, the quintessence
de-clustering parameter, $\zeta$, and the quintessence
equation of state parameter, $w$, with regard to partial
energy conservation.   Captions as in Fig.\,\ref{f:etam}.}
\label{f:etac}    
\end{figure}

The dependence, $\eta=\eta(m)$, is represented
in Fig.\,\ref{f:etac} with regard to the same
cases as in Fig.\,\ref{f:etap}, but different
degrees of quintessence de-clustering, $\zeta=0$
(squares), 0.25 (triangles), 0.5 (asterisks, 
0.75 (crosses), and 1 (diamonds).   The 
starting point on the left, $(0,0.5)$, marks
the limiting situation of vanishing quintessence,
and necessarily coincides in all cases.   It is
apparent that lower curves correspond to more
clustered quintessence and vice versa.   In
the limit, $w\to-1/3$, virialized configurations
coincide with their counterparts in absence of
quintessence, independent of the degree of 
clustering.   On the other hand, the dependence
is enhanced as the quintessence equation of
state parameter, $w$, attains lower values
up to $-1$.   In any case, the general trend
remains unchanged with respect to the special
situation of fully clustered quintessence.

In dealing with fully clustered quintessence,
an inspection of Figs.\,\ref{f:etam} and \ref
{f:etac} shows that partial energy conservation
yields larger virialized configurations, with
respect to their counterparts where global
energy conservation holds.   An opposite trend
is found in ML05, for $w=-4/5$, in the case of
fully clustered (Fig.\,2 therein) and unclustered
(Fig.\,4 therein) quintessence.   The reasons of
the above mentioned discrepancy are due to
different assumptions made in the current paper
and in ML05, as explained in Sect.\,\ref{cluq}.

\begin{figure}
\centerline{\psfig{file=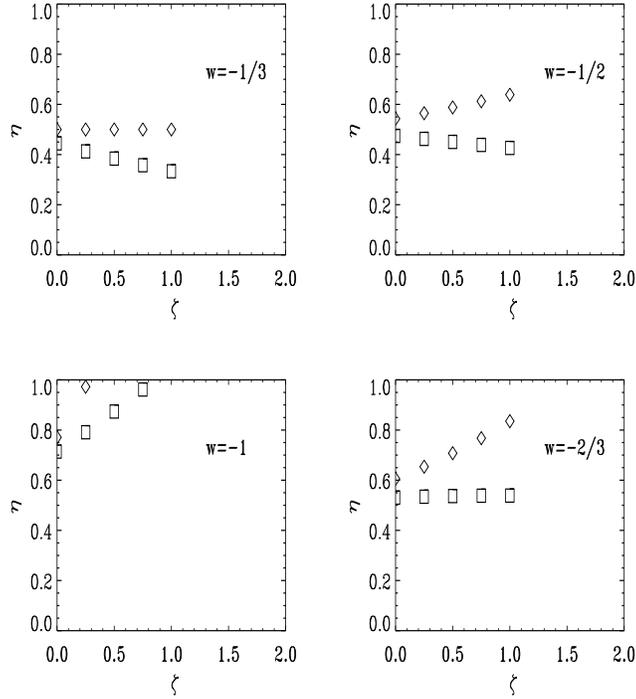,height=100mm,width=90mm}}
\caption{The fractional radius, $\eta=R_{\rm vir}/
R_{\rm max}$, as a function of the quintessence
de-clustering parameter, $\zeta$, for a fractional
mass, $m=M_q(R_{\rm max})/M_m$, and values of the quintessence
equation of state parameter, $w$, from top left in the
clockwise sense: $-$1/3, $-$1/2, $-$2/3, $-$1.
Global and partial energy conservation (in the
special case of fully clustered quintessence)
correspond to squares and diamonds, respectively.}
\label{f:etau}    
\end{figure}

The dependence, $\eta=\eta(\zeta)$, is represented
in Fig.\,\ref{f:etau} with regard to both global
(squares) and partial (diamonds) energy conservation
(in the special case of fully clustered quintessence),
for $m=1/4$ and $w=-1/3$, $-1/2$, $-2/3$, $-1$.
In both cases, larger fractional radii are 
attained for (algebraically) lower quintessence
equation of state parameters, $w$.   In addition,
with regard to fully clustered quintessence,
partial energy conservation yields larger
virialized configurations when compared to 
their counterparts where global energy
conservation holds.   Accordingly, partial
energy conservation implies a larger amount
of kinetic energy in matter subsystems, with
respect to global energy conservation.   An
opposite result has been found in ML05
(Fig.\,3 therein) for $w=-4/5$ and $m=1/5$.
The reasons of the above mentioned discrepancy
are due to different assumptions made in the
current paper and in ML05, as explained in
Sect.\,\ref{cluq}.

\begin{figure}
\centerline{\psfig{file=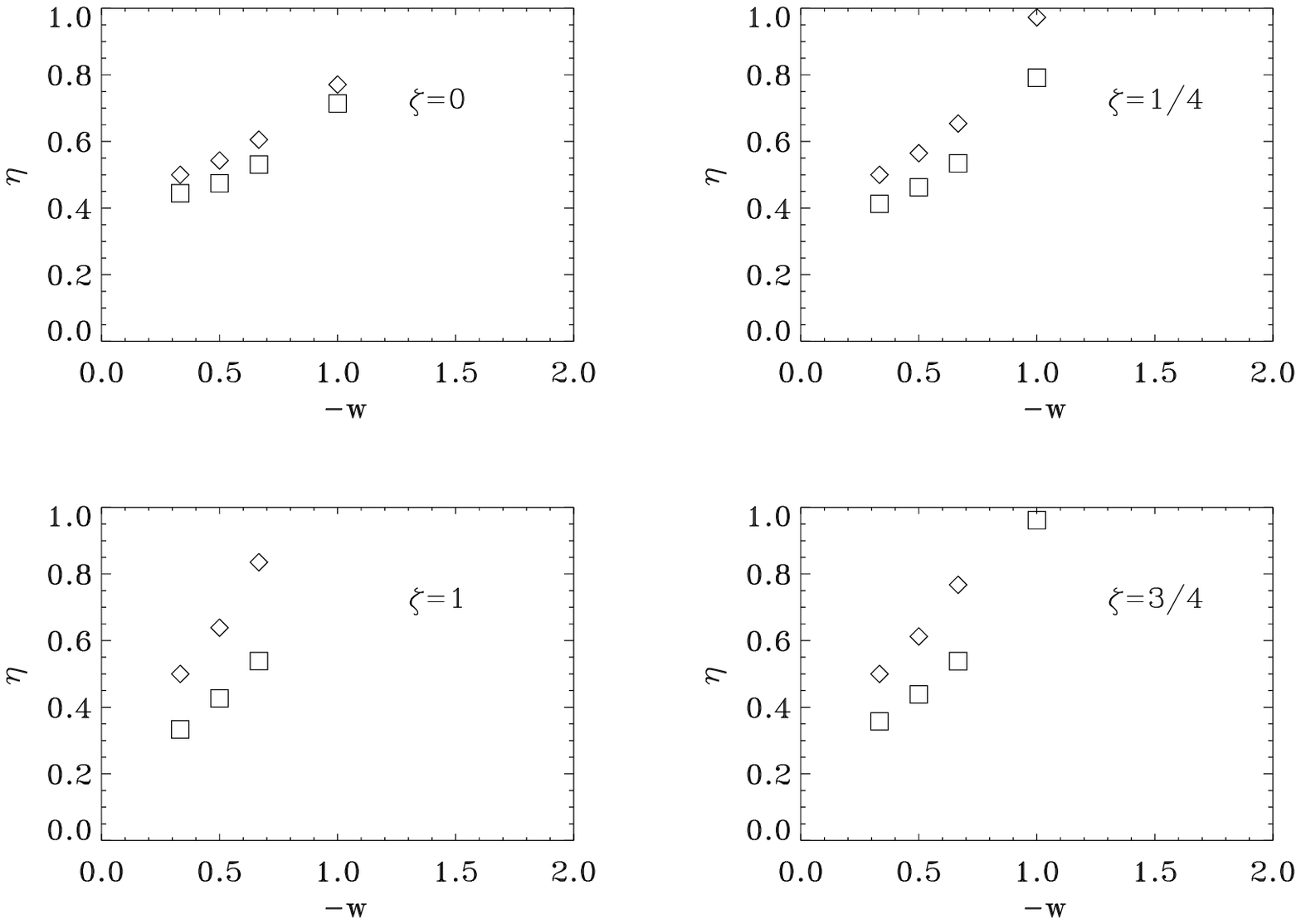,height=100mm,width=90mm}}
\caption{The fractional radius, $\eta=R_{\rm vir}/
R_{\rm max}$, as a function of the quintessence
equation of state parameter, $w$, for a fractional
mass, $m=1/4$, and values of the quintessence
de-clustering parameter, $\zeta$, from top left in the
clockwise sense: 0 (fully clustered quintessence),
1/4, 3/4, 1 (fully de-clustered quintessence).
Global and partial energy conservation (in the
special case of fully clustered quintessence)
correspond to squares and diamonds, respectively.}
\label{f:etaw}    
\end{figure}

The dependence, $\eta=\eta(w)$, is represented
in Fig.\,\ref{f:etaw} with regard to both global
(squares) and partial (diamonds) energy conservation
(in the special case of fully clustered quintessence),
for $m=1/4$ and $\zeta=0$ (fully clustered quintessence),
1/4, 3/4, 1 (fully de-clustered quintessence).
In both cases, the trend remains unchanged as
the degree of de-clustering gets increased.
Other features have already been discussed
above.   A different result has been found in
ML05 (Fig.\,5 therein), for $w=-4/5$ and
$m=1/5$.   The reasons of this discrepancy
are due to different assumptions made in the
current paper and in ML05, as explained in 
Sect.\,\ref{cluq}.

\section{Discussion.}
\label{disc}

In absence of a full knowledge on the dark
energy, an investigation on the virialization
of spherical overdensities appears to be
physically meaningful, even if particular
assumptions are made.   More specifically,
the potential induced by quintessence has
been conceived as due to quintessence
``particles'', interacting via an inverse
square distance law, with strength equal
to $(1+3w)G$, $-1\le w<-1/3$.   Accordingly,
the potential induced by a quintessence particle
scales as an inverse distance, ${\cal V}(r_p)
\propto1/r_p$, which makes the formulation of
the virial theorem, $2T=\kappa\Omega$,
${\cal V}(r_p)\propto r_p^\kappa$, $\kappa=-1$
(e.g., Landau \& Lifchitz 1966, Chap.\,II,
\S\,10).

With regard to the transition from
turnaround to virialization, the reference
case of energy conservation may be related
to either the whole (quintessence + matter)
system, or the matter subsystem within the
tidal field induced by the quintessence
subsystem.   According to ML05, global
energy conservation occurs in the special
case of fully clustered quintessence.   
The same is assumed with regard to partial
energy conservation, even if it is not
clearly specified in ML05 [Eq.\,(17)
therein].

Bearing in mind the reference cases
of global and partial energy conservtion,
different degrees of quintessence
clustering may be expressed via the
de-clustering parameter, $\zeta$, $0\le
\zeta\le1$, which is related to the
change in mass with respect to a
selected reference case, according
to Eqs.\,(\ref{eq:Dmu})-(\ref{eq:etaD})
and (\ref{eq:zita}).   The limiting
cases, $\zeta=0,1$, represent fully
clustered and fully de-clustered quintessence,
respectively.

Although typical values of the fractional
mass at turnaround, $m=$\linebreak
$M_q(T_{\rm max})/M_m$,
do not exceed a value of about 0.9 (e.g.,
ML05), the analysis of a larger range
shows interesting features, namely (i)
the existence of a threshold in $m$,
increasing with the quintessence
equation of state parameter, $w$,
above which no virialized configuration
is allowed, see Figs.\,\ref{f:etam} and
\ref{f:etac}, and (ii) the occurrence
of a non monotonic trend in the fractional
radius, $\eta$, as a function of the
fractional mass, $m$, with the occurrence
of a minimum, for $w=-1/2$, a decreasing
monotonic trend for $w=-1/3$, and an
increasing monotonic trend for $w=-2/3,
-1$, with regard to global energy
conservation in the limit of fully
clustered quintessence, see Fig.\,\ref
{f:etam}.

The dependence, $\eta=\eta(\zeta)$, related
to $m=1/4$, is also monotonic for both
global and partial energy conservation
in the limiting case of fully clustered
quintessence, see Fig.\,\ref{f:etau}.
Owing to Eq.\,(\ref{eq:etau}), the
special case of unclustered quintessence
occurs within the whole range, $0\le\zeta
<1$, according if $1\le a_{\rm vir}/a_
{\rm max}<1500$ and/or $-1\le w<-1/3$.

The dependence, $\eta=\eta(w)$, shows a
monotonic trend for both global and
partial energy conservation
in the limiting case of fully clustered
quintessence, see Fig.\,\ref{f:etaw},
with increasingly different trends as
$\zeta$ increases.   It can be seen
that virialized configurations cannot
take place for both sufficiently low $w$
and sufficiently large $\zeta$.

In any case, more extended virialized
configurations occur for partial (with
respect to global) energy conservation
in the limiting case of fully clustered
quintessence, see Figs.\,\ref{f:etau}
and \ref{f:etaw}.   Accordingly, the
latter alternative yields bounder
configurations with respect to the
former.   In addition, an increasing
degree of quintessence de-clustering
makes more extended virialized
configurations provided partial
energy conservation holds in the
special case of fully clustered
quintessence, see Fig.\,\ref{f:etau}.
When global energy conservation
holds in the special case of fully
clustered quintessence, a similar
trend occurs for sufficiently low
quintessence equation of state
parameters, $w<-2/3$.   No change
appears for $w=-2/3$, while the
trend is reversed for $w>-2/3$.

The above results are different
from their analogon in ML05,
even if the expression of the
potential energy is the same
as in the current paper.   The
discrepancy could be due to
(i) different formulations of
the virial theorem, as discussed
in Sect.\,\ref{cluq} and in
Appendix A, with regard to both
the whole system and the matter
subsystem, and (ii) different
descriptions of de-clustered
quintessence, by defining a
quintessence de-clustering
parameter in the current paper,
and modifying the quintessence
continuity equation in ML05,
Eqs.\,(9)-(11) therein.

If the fractional radius, $\eta$,
can be deduced from observations,
it (together with the related
uncertainty) translates into the
plane of Figs.\,\ref{f:etam}-\ref
{f:etaw} as a horizontal band,
and the value of the fractional
mass, $m$, the degree of quintessence
de-clustering, $\zeta$, and the
quintessence equation of state
parameter, $w$, can be constrained.
In addition, global and partial
energy conservation in the special
case of fully clustered quintessence,
can be tested.

\section{Conclusion}\label{conc}

The virialization of matter overdensities
within dark energy subsystems has been
considered under a number of simplifying
assumptions, namely (i) spherical-symmetric
density profiles, (ii) time-independent
quintessence equation of state parameter,
$w$, and (iii) sole gravitational interaction
between dark energy scalar field and matter.
The quintessence subsystem has been conceived
as made of ``particles'' whose mutual interaction
has intensity equal to $(1+3w)G$ and scales as
the inverse square of the distance.   The related
expression of self and tidal potential energy and
formulation of the virial theorem for subsystems,
have been found to be consistent with their matter
counterparts, passing from $-1\le w<-1/3$ to $w=0$.
In the special case of fully clustered quintessence,
following ML05, it has been assumed either global
energy conservation related to the whole system,
or partial energy conservation related to the matter
subsystem within the tidal field induced by the
quintessence subsystem.   Further investigation
has been devoted to a few special cases, namely
a limiting situation, $w=-1/3$, and lower values,
$w=-1/2, -2/3, -1$, where the last one mimics the
presence of a cosmological constant.

The special case of fully clustered (i.e.
collapsing together with the matter) quintessence
has been studied in detail, following a similar
procedure as in ML05.   The general case of
partially clustered quintessence has been
considered in terms of a degree of quintessence
de-clustering, $\zeta$, $0\le\zeta\le1$, ranging
from fully clustered $(\zeta=0)$ to completely
de-clustered $(\zeta=1)$ quintessence, respectively.
The special case of unclustered (i.e. remaining
homogeneous) quintessence has also been discussed.

The trend exhibited by the fractional radius,
$\eta$, as a function of the fractional mass,
$m$, the degree of quintessence de-clustering,
$\zeta$, and the quintessence equation of state
parameter, $w$, has been found to be different
from its counterpart reported in earlier attempts
(e.g., ML05).   In particular, no clear dichotomy
with respect to the limiting situation of a
vanishing quintessence, $\eta=1/2$, has been
shown when global or partial energy conservation
holds in the special case of fully clustered
quintessence, with $\eta>1/2$ preferred.   The
reasons of the above mentioned discrepancy have
been recognized as owing to (i) different
formulations of the virial theorem, and (ii)
different descriptions of de-clustered quintessence,
with respect to the case of fully clustered
quintessence.


\section*{Appendix}

\subsection*{A. On the virial theorem for subsystems}
\label{vits}

Let $i$ and $j$ denote concentric, spherical-symmetric,
homogeneous subsystems, $R_i$ and $R_j$, $R_i\le R_j$,
the related radii, and let the potential and the
potential energy terms be expressed by Eqs.\,(\ref
{eq:pote}), (\ref{eq:sos}), (\ref{eq:soii}), (\ref
{eq:soti}), and (\ref{eq:sotjB}).   The derivation
of the potential energy terms with respect to the
inner or outer radius yields:
\begin{lefteqnarray}
\label{eq:dOu}
&& R_i\frac{\partial\Omega_i}{\partial R_i}=-
\Omega_i~~;\qquad R_j\frac{\partial\Omega_i}
{\partial R_j}=0~~; \\
\label{eq:dOv}
&& R_j\frac{\partial\Omega_j}{\partial R_j}=-
\Omega_j~~;\qquad R_i\frac{\partial\Omega_j}
{\partial R_i}=0~~; \\
\label{eq:dWij} 
&& \frac{R_i}{1+3w_j}\frac{\partial W_{ij}}{\partial R_i}=
\frac{R_i}{1+3w_i}\frac{\partial W_{ji}}{\partial R_i}=
-\frac12\frac{V_{ij}}{1+3w_j}~~; \\                                
\label{eq:dWji} 
&& \frac{R_j}{1+3w_i}\frac{\partial W_{ji}}{\partial R_j}=
\frac{R_j}{1+3w_j}\frac{\partial W_{ij}}{\partial R_j}=
-\frac{V_{ij}}{1+3w_j}\left(\frac54y^2-\frac34\right)~~; \\                                
\label{eq:dVij}
&& R_i\frac{\partial V_{ij}}{\partial R_i}=2V_{ij}~~;\qquad
R_j\frac{\partial V_{ij}}{\partial R_j}=-3V_{ij}~~; \\
\label{eq:dViji}
&& R_j\frac{\partial V_{ji}}{\partial R_j}=-V_{ij}
\left[\left(\frac54y^2-\frac{15}4\right)+\left(\frac54
y^2-\frac34\right)\frac{1+3w_i}{1+3w_j}\right]~~; \\
\label{eq:dVjii}
&& R_i\frac{\partial V_{ji}}{\partial R_i}=-V_{ij}
\left(\frac52+\frac12\frac{1+3w_i}{1+3w_j}\right)~~;
\end{lefteqnarray}
which, in turn, produce:
\begin{lefteqnarray}
\label{eq:sdOu}
&& R_u\frac{\partial\Omega_u}{\partial R_u}+
R_v\frac{\partial\Omega_u}{\partial R_v}=-
\Omega_u~~;\qquad u=i,j~~;\qquad v=j,i~~; \\
\label{eq:sdVuv}
&& R_u\frac{\partial V_{uv}}{\partial R_u}+
R_v\frac{\partial V_{uv}}{\partial R_v}=-V_{uv}~~;
\qquad u=i,j~~;\qquad v=j,i~~;
\end{lefteqnarray}
owing to Eqs.\,(\ref{eq:viru}), (\ref{eq:sdOu}),
and (\ref{eq:sdVuv}), the virial theorem for
subsystems may be formulated as:
\begin{lefteqnarray}
\label{eq:dviru}
&& 2T_u-R_u\frac{\partial\Omega_u}{\partial R_u}-
R_v\frac{\partial\Omega_u}{\partial R_v}-
R_u\frac{\partial V_{uv}}{\partial R_u}-
R_v\frac{\partial V_{uv}}{\partial R_v}=0~~;
\nonumber \\
&& u=i,j~~;\qquad v=j,i~~;
\end{lefteqnarray}
and the sum of the two alternative expressions
reads:
\begin{leftsubeqnarray}
\slabel{eq:dvira}
&& 2T-R_i\frac{\partial\Omega}{\partial R_i}-
R_j\frac{\partial\Omega}{\partial R_j}=0~~; \\
\slabel{eq:dvirb}
&& T=T_i+T_j~~; \\
\slabel{eq:dvirc}
&& \Omega=\Omega_i+V_{ij}+V_{ji}+\Omega_j=
\Omega_i+W_{ij}+W_{ji}+\Omega_j~~;
\label{seq:dvir}
\end{leftsubeqnarray}
where $T$ and $\Omega$,
according to Eq.\,(\ref{eq:VW}),
are the kinetic and the potential energy,
respectively, of the whole system.

The explicit expression of the last two
terms on the left-hand side of Eq.\,(\ref
{eq:dvira}), and the related sum, may be
calculated using Eqs.\,(\ref{eq:dOu})-(\ref
{eq:dVjii}).   The result is:
\begin{leftsubeqnarray}
\slabel{eq:dOa}
&& R_i\frac{\partial\Omega}{\partial R_i}=-
\Omega_i-\frac12\left(1+\frac{1+3w_i}{1+3w_j}
\right)V_{ij}~~; \\
\slabel{eq:dOb}
&& R_j\frac{\partial\Omega}{\partial R_j}=-
\Omega_j-\left(\frac54y^2-\frac34\right)\left(1+\frac{1+
3w_i}{1+3w_j}\right)V_{ij}~~; \\
\slabel{eq:dOc} 
&& R_i\frac{\partial\Omega}{\partial R_i}+
R_j\frac{\partial\Omega}{\partial R_j}=-         
\Omega_i-\Omega_j-\left(\frac54y^2-\frac14\right)
\left(1+\frac{1+3w_i}{1+3w_j}\right)V_{ij}~~;
\label{seq:dVjii}
\end{leftsubeqnarray}
in the special case where the two subsystems
fill the same volume, $R_i=R_j=R$ or $y=1$,
Eq.\,(\ref{eq:Ot}) holds and Eqs.\,(\ref
{seq:dvir}) reduce to:
\begin{leftsubeqnarray}
\slabel{eq:dOla}
&& \lim_{R_i\to R}\left(R_i\frac{\partial
\Omega}{\partial R_i}\right)=-
\Omega_i-\frac12V_{ij}-\frac12V_{ji}~~; \\
\slabel{eq:dOlb}
&& \lim_{R_j\to R}\left(R_j\frac{\partial
\Omega}{\partial R_j}\right)=-
\Omega_j-\frac12V_{ji}-\frac12V_{ij}~~; \\
\slabel{eq:dOlc}
&&\lim_{R_i\to R}\left(R_i\frac{\partial
\Omega}{\partial R_i}\right)+\lim_{R_j\to R}
\left(R_j\frac{\partial\Omega}{\partial R_j}
\right)=-\Omega_i-V_{ij}-V_{ji}-\Omega_i=
-\Omega~~;
\label{seq:dOl}
\end{leftsubeqnarray}
in conclusion, the total potential
energy of the whole system must be
conceived as dependent on four independent
variables, $\Omega=\Omega(M_i, M_j, R_i,
R_j)$.

On the other hand, one-component systems
are subjected to no tidal potential, then
the potential energy coincides with the
self potential energy, which depends on
two independent variables, $\Omega=\Omega
(M, R)$.   Accordingly,  Eq.\,(\ref{eq:dvira})
translates into:
\begin{equation}
\label{eq:dvirt}
2T-R\frac{\partial\Omega}{\partial R}=0~~;
\end{equation}
which cannot be splitted as a sum of different
contributions, as done in e.g., ML05, unless it
is conceived
as a function of four independent variables
instead of two, and Eq.\,(\ref{eq:dvira}) is
used instead of Eq.\,(\ref{eq:dvirt}).

\subsection*{B. The function $f(x)$}
\label{fx}

Let us define the function:
\begin{leftsubeqnarray} 
\slabel{eq:fa}
&& f(x)=\frac{1+x}{1+m}~\frac{1+(1+3w)x}{1+(1+3w)m}~~; \\
\slabel{eq:fb}
&& 0\le x\le m~~;\qquad0\le f(x)\le2~~;\qquad
-1\le w<-\frac13~~;
\label{seq:f}
\end{leftsubeqnarray}
where the values at the extrema of the
domain are:
\begin{equation}
\label{eq:f0}
f(0)=\frac1{1+(2+3w)m+(1+3w)m^2}~~;\qquad f(m)=1~~;
\end{equation}
and keeping in mind that the equation:
\begin{equation}
\label{eq:df0}
(1+x)[1+(1+3w)x]=0~~;
\end{equation}
admits the real solutions:
\begin{equation}
\label{eq:sdf0}
x_1=-1~~;\qquad x_2=-\frac1{1+3w}~~;
\end{equation}
the sign of the function, $f(x)$, is
determined by the inequalities:
\begin{leftsubeqnarray} 
\slabel{eq:sfa}
&& f(x)>0~~;\qquad0\le m<\frac{-1}{1+3w}~~;\qquad
0\le x<\frac{-1}{1+3w}~~; \\
\slabel{eq:sfb}
&& f(x)<0~~;\qquad m>\frac{-1}{1+3w}~~;\qquad
0\le x<\frac{-1}{1+3w}~~; \\
\slabel{eq:sfc}
&& f(x)>0~~;\qquad m>\frac{-1}{1+3w}~~;\qquad
\frac{-1}{1+3w}<x\le m~~;
\label{seq:sf}
\end{leftsubeqnarray}
and the function is null at $x=-1$
and $x=-1/(1+3w)$, respectively.   In the
special cases, $m=-1$ and $m=-1/(1+3w)$,
the function diverges everywhere within the
domain, except at $x=m$ where $f(m)=1$.

The first and the second derivatives are:
\begin{lefteqnarray}
\label{eq:d1f}
&& \frac{\diff f}{\diff x}=\frac{(2+3w)+2
(1+3w)x}{1+(2+3w)m+(1+3w)m^2}~~; \\
\label{eq:d2f}
&& \frac{\diff^2f}{\diff x^2}=\frac{2
(1+3w)}{1+(2+3w)m+(1+3w)m^2}~~;
\end{lefteqnarray}
where the first derivative is null
at the extremum point:
\begin{equation}
\label{eq:xe}
x^\dagger=-\frac12\frac{2+3w}{1+3w}~~;
\end{equation}
and the sign of the second derivative
is defined by the inequalities:
\begin{leftsubeqnarray} 
\slabel{eq:sxea}
&& \frac{\diff^2f}{\diff x^2}<0~~;\qquad
0\le m<\frac{-1}{1+3w}~~; \\
\slabel{eq:sxeb}
&& \frac{\diff^2f}{\diff x^2}>0~~;\qquad
m>\frac{-1}{1+3w}~~;
\label{seq:sxe}
\end{leftsubeqnarray}
in the range of interest, $-1\le w<-1/3$.
Accordingly, the extremum point, $x^\dagger$,
is a maximum and a minimum, respectively.

At the extremum point, the function attains
the value:
\begin{equation}
\label{eq:fxe}
f(x^\dagger)=-\frac94\frac{w^2}{1+3w}\frac1
{1+(2+3w)m+(1+3w)m^2}~~;
\end{equation}
on the other hand, the extremum point is
attained at the upper limit of the domain,
$x^\dagger=m$, provided the parameter, $w$,
has the value:
\begin{equation}
\label{eq:wlm}
w=-\frac23\frac{m+1}{2m+1}~~;
\end{equation}
in particular, $w=-2/3$ for $x^\dagger=m=0$
and $w=-1/3$ for $x^\dagger=m\to+\infty$.
Within the range, $-1\le w<-2/3$, the extremum
point lies outside the domain, $x^\dagger<0$.

To gain further insight, a few special cases
shall be studied with more detail, namely
$w=-1,\, -2/3,\, -1/2,\, -1/3$, the last to
be conceived as an interesting limiting
situation.   The general case, $-1\le w<-1/3$,
is expected to show similar properties as
in the closest among the above mentioned
special cases.

\begin{figure}
\centerline{\psfig{file=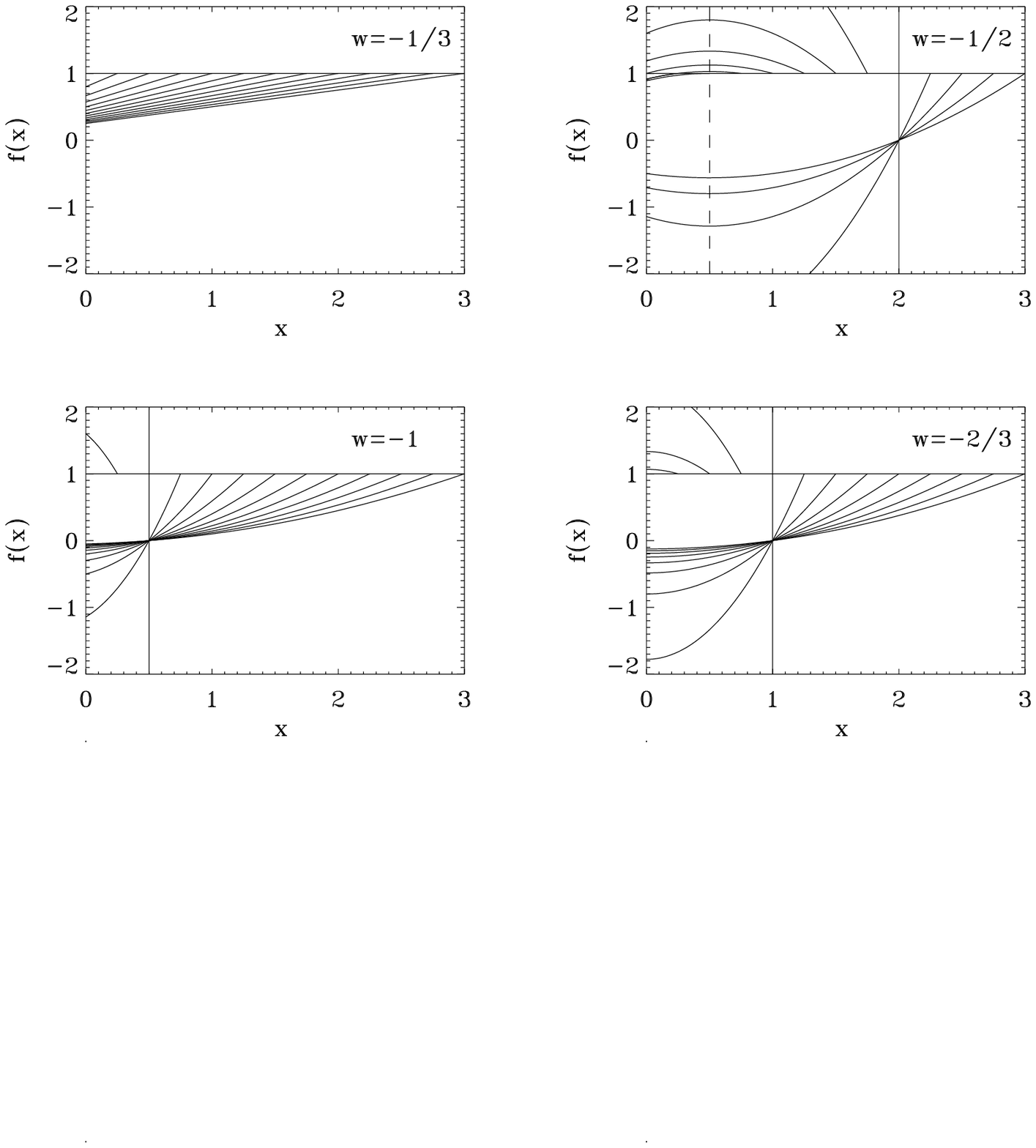,height=100mm,width=90mm}}
\caption{The function, $f(x)$, for different
values of the parameter, $m$, within the
domain, $0\le x\le m$, with regard to
special choices of the parameter, $w$,
(from top left in the clockwise sense):
$-1/3,\, -1/2,\, -2/3,\, -1$.   The value
of $m$ related to each curve, starting from 
the left, is $m=i/4$, $0\le i\le12$.   The
horisontal axis corresponds to $m\to+\infty$.
The locus of ending points is the horisontal
line, $f(x)=1$.   In the special case,
$m=0$, the curve reduces to a single point,
$(0,1)$.   The locus of extremum points is
marked by a dashed vertical line for $w=
-1/2$, and coincides with the vertical axis
for $w=-2/3$.}
\label{f:b1}    
\end{figure}

In the special case, $w=-1/3$, Eq.\,(\ref
{eq:fa}) reduces to:
\begin{equation}
\label{eq:f1}
f(x)=\frac{1+x}{1+m}~~;
\end{equation}
which is positive within the domain, 
$0\le x\le m$, according to Eq.\,(\ref
{eq:sfa}).   The extremum point is a
maximum and occurs at infinite,
according to Eqs.\,(\ref{eq:sxea}) and
(\ref{eq:fxe}).   Then it belongs to
the domain only in the limit, $m\to+
\infty$.   The function, $f(x)$, for
different values of the parameter, $m$,
is represented in Fig.\,\ref{f:b1}, top
left.

In the special case, $w=-1/2$, Eq.\,(\ref
{eq:fa}) reduces to:
\begin{equation}
\label{eq:f2}
f(x)=\frac{2+x-x^2}{2+m-m^2}~~;
\end{equation}
which, according to  Eqs.\,(\ref{seq:sf}),
is positive within the domain, $0\le x\le
m$, for $m<2$, and within the domain,
$2<x\le m$, for $m>2$; on the other hand,
it is negative within the domain, $0\le
x<2$, for $m>2$.   The (allowed) zero of
the function occurs at $x=2$.   In the
special case, $m=2$, the function diverges
everywhere within the domain, except at
$x=2$, where $f(2)=1$.

Owing to Eq.\,(\ref{eq:xe}), the extremum
point occurs at $x^\dagger=1/2$, which is
a maximum for $0\le m<2$ and a minimum
for $m>2$, according to Eqs.\,(\ref{seq:sxe}).
The value of the function at the extremum
point is:
\begin{equation}
\label{eq:ef2}
f(x^\dagger)=\frac94\frac1{2+m-m^2}~~;\qquad
x^\dagger=\frac12~~;
\end{equation}
which belongs to the domain only if $m\ge1/2$,
$m\ne2$.   The function, $f(x)$, for
different values of the parameter, $m$,
is represented in Fig.\,\ref{f:b1}, top
right.

In the special case, $w=-2/3$, Eq.\,(\ref
{eq:fa}) reduces to:
\begin{equation}
\label{eq:f3}
f(x)=\frac{1-x^2}{1-m^2}~~;
\end{equation}
which, according to  Eq.\,(\ref{seq:sf}),
is positive within the domain, $0\le x\le
m$, for $m<1$, and within the domain,
$1<x\le m$, for $m>1$; on the other hand,
it is negative within the domain, $0\le
x<1$, for $m>1$.   The (allowed) zero of
the function occurs at $x=1$.   In the
special case, $m=1$, the function diverges
everywhere within the domain, except at
$x=1$, where $f(1)=1$.

Owing to Eq.\,(\ref{eq:xe}), the extremum
point occurs at $x^\dagger=0$, which is
a maximum for $0\le m<1$ and a minimum
for $m>1$, according to Eqs.\,(\ref{seq:sxe}).
The value of the function at the extremum
point is:
\begin{equation}
\label{eq:ef3}
f(x^\dagger)=\frac1{1-m^2}~~;\qquad
x^\dagger=0~~;
\end{equation}
which, in any case, belongs to the domain.
The function, $f(x)$, for
different values of the parameter, $m$,
is represented in Fig.\,\ref{f:b1},
bottom right.

In the special case, $w=-1$, Eq.\,(\ref
{eq:fa}) reduces to:
\begin{equation}
\label{eq:f4}
f(x)=\frac{1-x-2x^2}{1-m-2m^2}~~;
\end{equation}
which, according to  Eqs.\,(\ref{seq:sf}),
is positive within the domain, $0\le x\le
m$, for $m<1/2$, and within the domain,
$1/2<x\le m$, for $m>1/2$; on the other hand,
it is negative within the domain, $0\le
x<1/2$, for $m>1/2$.   The (allowed) zero of
the function occurs at $x=1/2$.   In the
special case, $m=1/2$, the function diverges
everywhere within the domain, except at
$x=1/2$, where $f(1/2)=1$.

Owing to Eq.\,(\ref{eq:xe}), the extremum
point occurs at $x^\dagger=-1/4$, which is
outside the domain.   The function, $f(x)$,
for different values of the parameter, $m$,
is represented in Fig.\,\ref{f:b1}, bottom
left.

\subsection*{C. The effect of partially
clustered quintessence on the virialized
configuration}\label{pcqv}

With regard to a virialized configuration
in the special case of fully clustered
quintessence, let $\Delta\Omega$ be the
energy change, $E-E_{\rm FC}$, which
defines the virialized
configuration in the case of partially
clustered quintessence, related to
$\Delta\mu=\zeta\eta_{\rm FC}^{-3w}$,
according to Eqs.\,(\ref{eq:Dmu})-(\ref
{eq:etaD}), (\ref{eq:DOv}), and (\ref
{eq:zita}).   The combination of 
Eqs.\,(\ref{eq:muD}), (\ref{eq:DOv}),
and (\ref{eq:zita}) yields:
\begin{leftsubeqnarray} 
\slabel{eq:DOma}
&& \Delta\Omega=-\frac35\frac{GM_m^2}{R_{\rm max}}
\frac1{\eta_{\rm FC}}\left[(1+3w)\zeta(2-\zeta)x^2
+(2+3w)\zeta x\right]~~; \\
\slabel{eq:DOmb}
&& x=m\eta_{\rm FC}^{-3w}~~;\qquad0\le x\le m~~;
\label{seq:DOm}
\end{leftsubeqnarray}
and the substitution of Eq.\,(\ref{eq:DOmb})
into (\ref{eq:phi}) produces:
\begin{equation}
\label{eq:phx}
\phi(x;w,\zeta)=(1+3w)\zeta(2-\zeta)x^2+(2+3w)
\zeta x~~;
\end{equation}
which makes Eq.\,(\ref{eq:DOma}) be cast into the
form:
\begin{equation}
\label{eq:DOp}
\Delta\Omega=-\frac35\frac{GM_m^2}{R_{\rm max}}
\frac1{\eta_{\rm FC}}\phi~~;
\end{equation}
accordingly, the sign of the energy change,
$\Delta\Omega$, is opposite to the sign of
the function, $\phi(x)$.   Keeping in mind
that positive and negative $\Delta\Omega$
imply expansion and contraction, respectively,
with regard to the virialized configuration
in the special case of fully clustered
quintessence, the effect of partial clustering
may be deduced from the sign of $\phi(x)$.
According to Eq.\,(\ref{eq:phx}), the
solutions of $\phi(x)=0$ are:
\begin{equation}
\label{eq:phz}
x_1=0~~;\qquad x_2=-\frac1{2-\zeta}\frac{2+3w}
{1+3w}~~;
\end{equation}
and the sign of $\phi(x)$ is defined as:
\begin{leftsubeqnarray} 
\slabel{eq:sgpa}
&& \phi(x)\ge0~~;\qquad\min(x_1,x_2)\le x\le\max
(x_1,x_2)~~; \\
\slabel{eq:sgpb}
&& \phi(x)\le0~~;\qquad x\le\min(x_1,x_2)~~;
\qquad x\ge\max(x_1,x_2)~~;
\label{seq:sgp}
\end{leftsubeqnarray}
where $0\le x\le m$ in the case under discussion,
conform to Eq.\,(\ref{seq:f}).

The first derivative:
\begin{equation}
\label{eq:php}
\frac{\diff\phi}{\diff x}=2\zeta(2-\zeta)(1+3w)
x+\zeta(2+3w)~~;
\end{equation}
is null at the abscissa:
\begin{equation}
\label{eq:xmx}
x^\dagger=-\frac12\frac1{2-\zeta}\frac{2+3w}
{1+3w}~~;
\end{equation}
where, owing to Eq.\,(\ref{eq:phz}),
$\vert x^\dagger-x_1\vert=\vert x^
\dagger-x_2\vert$, or $x^\dagger=x_2/2$.

The second derivative:
\begin{equation}
\label{eq:phs}
\frac{\diff^2\phi}{\diff x^2}=2\zeta
(2-\zeta)(1+3w)~~;
\end{equation}
is negative within the range of interest,
$-1\le w<-1/3$, $0\le\zeta\le1$, which
implies that the extremum point is a
maximum.   The substitution of Eq.\,(\ref
{eq:xmx}) into (\ref{eq:phx}) allows the
calculation of the function at the maximum.
The result is:
\begin{equation}
\label{eq:phm}
\phi(x^\dagger)=-\frac14\frac\zeta{2-\zeta}
\frac{(2+3w)^2}{1+3w}~~;
\end{equation}
which is non negative, $\phi(x^\dagger)\ge0$,
in the case under discussion.

Further inspection of Eq.\,(\ref{eq:xmx})
shows that:
\begin{leftsubeqnarray} 
\slabel{eq:xmsa}
&& x^\dagger>0~~;\qquad-\frac23<w<-\frac13~~; \\
\slabel{eq:xmsb}
&& x^\dagger<0~~;\qquad-1<w<-\frac23~~; \\
\slabel{eq:xmsc}
&& x^\dagger=0~~;\qquad w=-\frac23~~; \\
\slabel{eq:xmsd}
&& x^\dagger\to+\infty~~;\qquad w\to\left(-\frac13
\right)^-~~;
\label{seq:xmsc}
\end{leftsubeqnarray}
independent of the value of $\zeta$
within the assigned range.

The particularization of Eqs.\,(\ref{eq:phx}),
(\ref{eq:phz}), (\ref{eq:xmx}), and (\ref
{eq:phm}), to a few special values of
$\zeta$, yields:
\begin{leftsubeqnarray} 
\slabel{eq:p13a}
&& \phi(x)=\zeta x~~; \\
\slabel{eq:p13b}
&& \phi(x^\dagger)\to+\infty~~;\qquad x^\dagger
=\frac12x_2\to+\infty~~;
\label{seq:p13}
\end{leftsubeqnarray}
in the special case, $w=-1/3$;
\begin{leftsubeqnarray} 
\slabel{eq:p12a}
&& \phi(x)=-\frac12\zeta\left[(2-\zeta)x^2-x
\right]~~; \\
\slabel{eq:p12b}
&& \phi(x^\dagger)=\frac18\frac\zeta{2-\zeta}~~;
\qquad x^\dagger=\frac12x_2=\frac12\frac1
{2-\zeta}~~;
\label{seq:p12}
\end{leftsubeqnarray}
in the special case, $w=-1/2$;
\begin{leftsubeqnarray} 
\slabel{eq:p23a}
&& \phi(x)=-\zeta(2-\zeta)x^2~~; \\
\slabel{eq:p23b}
&& \phi(x^\dagger)=0~~;\qquad x^\dagger
=\frac12x_2=0~~;
\label{seq:p23}
\end{leftsubeqnarray}
in the special case, $w=-2/3$;
\begin{leftsubeqnarray} 
\slabel{eq:p10a}
&& \phi(x)=-\zeta\left[2(2-\zeta)x^2+x\right]~~; \\
\slabel{eq:p10b}
&& \phi(x^\dagger)=\frac18\frac\zeta{2-\zeta}~~;
\qquad x^\dagger=\frac12x_2=-\frac14\frac1{2-\zeta}~~;
\label{seq:p10}
\end{leftsubeqnarray}
in the special case, $w=-1$.

\begin{figure}
\centerline{\psfig{file=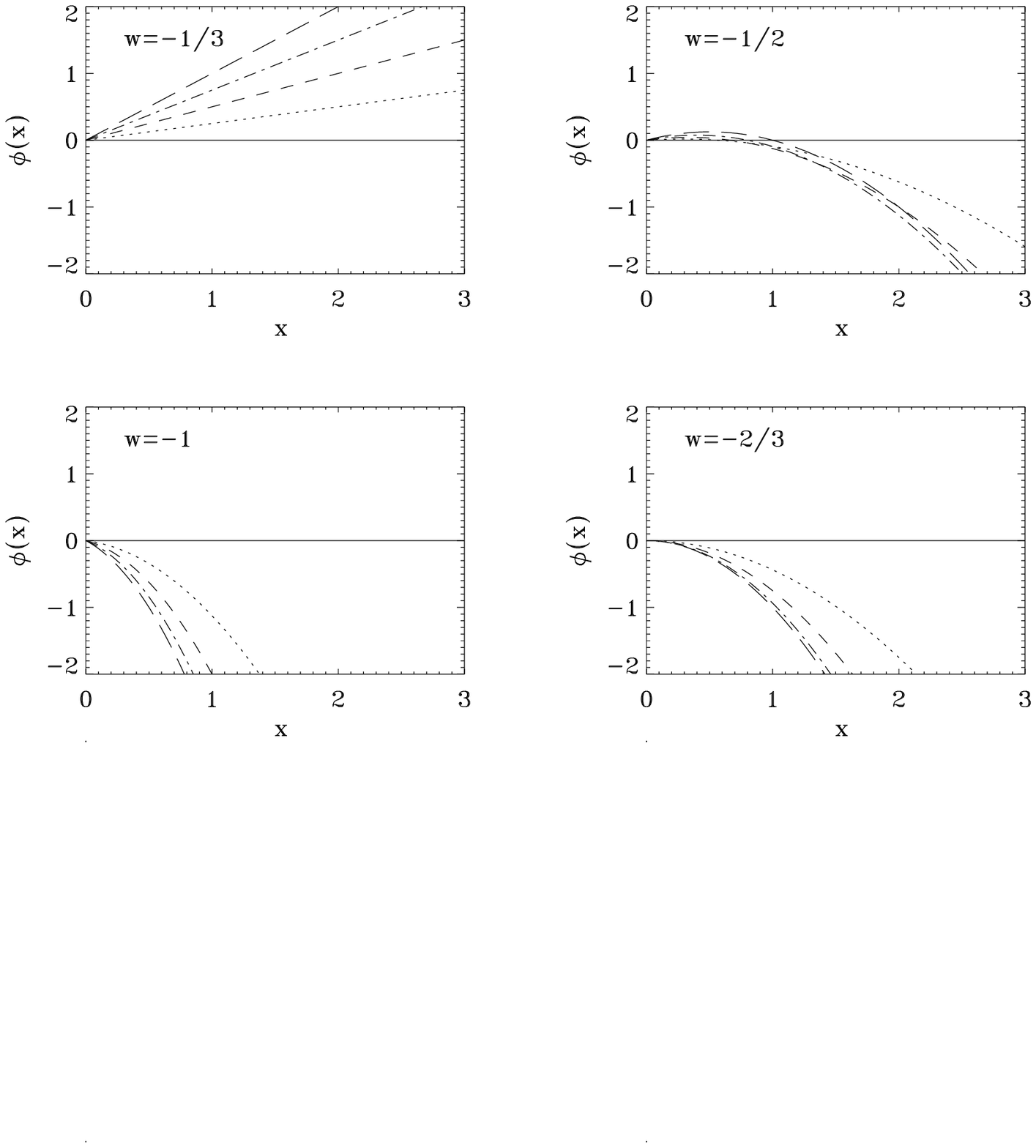,height=100mm,width=90mm}}
\caption{The function, $\phi(x)$, for different
values of the parameter, $\zeta$, with regard to
special choices of the parameter, $w$,
(from top left in the clockwise sense):
$-1/3,\, -1/2,\, -2/3,\, -1$.   Dotted
curves correspond to $\zeta=1/4$, dashed
curves to $\zeta=1/2$, dot-dashed curves
to $\zeta=3/4$, and long-dashed curves to
$\zeta=1$, which is the limit of fully
de-clustered quintessence.   In the limit
of fully clustered quintessence, $\zeta
=0$, all curves coincide with the horisontal
axis.}
\label{f:c1}    
\end{figure}

The function,
$\phi(x)$, expressed by Eqs.\,(\ref
{eq:p13a})-(\ref{eq:p10a}) and related to
different values of $\zeta$, are represented
in Fig.\,\ref{f:c1} with regard to the special
cases considered above.
                                
Keeping in mind Eq.\,(\ref{eq:DOp}),
the transition from a virialized
configuration related to fully clustered
quintessence to its counterpart related
to partially clustered quintessence,
implies expansion $(\phi<0)$ for values
of the quintessence equation of state
parameter, $w$, in the range, $-1\le
w<-2/3$.   On the other hand, both
expansion and contraction $(\phi>0)$
may occur for $-2/3\le w<-1/3$,
according if the variable, $x$, is
sufficiently distant from 0 and/or
the parameter, $w$, is sufficiently
close to $-2/3$, and vice versa.


\begin{thebibliography}{}
\bibitem{} Amendola, L. 2000, PhRvD 62, 043511
\bibitem{} Battye, R.A., Weller, J. 2003, PhRvD 68, 083506
\bibitem{} Brosche, P., Caimmi, R., Secco, L. 1983, A\&A 125, 338
\bibitem{} Caimmi, R. 2003, AN 324, 250
\bibitem{} Caimmi, R., Secco, L., Brosche, P. 1984, A\&A 139, 411
\bibitem{} Caimmi, R., Secco, L. 1992, ApJ 395, 119
\bibitem{} Caldwell, R.R., Dave, R., Steinhardt, P.J. 1988a, ApSS 261, 303
\bibitem{} Caldwell, R.R., Dave, R., Steinhardt, P.J. 1988b, PhRvL 80, 1582
\bibitem{} Chandrasekhar, S. 1969, Ellipsoidal Figures
             of Equilibrium, Yale University Press, New Haven
\bibitem{} Gunn, J.E., Gott, J.R., III 1972, ApJ 176, 1
\bibitem{} Horellou, C., Berge, J. 2005, MNRAS 360, 1393
\bibitem{} Iliev, I.T., Shapiro, P.R. 2001, MNRAS 325, 468 
\bibitem{} Lahav, O., Lilje, P.B., Primack, J.R., Rees, M.J. 1991,
           MNRAS 251, 128
\bibitem{} Landau, L., Lifchitz, E. 1966 {\it Mecanique}, Mir, Moscow
\bibitem{} Limber, D.N. 1959, ApJ 130, 414
\bibitem{} MacMillan, D.W. 1930 {\it The Theory of the Potential},
           Dover Publications, Inc. New York, 1958
\bibitem{} Maor, I., Lahav, O. 2005, JCAP 7, 3 (ML05)
\bibitem{} Mota, D.F., van de Bruck, C. 2004, A\&A 421, 71
\bibitem{} Peebles, P.J.E. Ratra, B. 1988, ApJ 325, 17
\bibitem{} Peebles, P.J.E. Ratra, B. 2003, Rev. Mod. Phys. 75, 559
\bibitem{} Percival, W.J. 2005, A\&A 443, 819
\bibitem{} Ratra, B. 1988, Peebles, P.J.E. 1988, PhRvD 37, 3406
\bibitem{} Rubi\~no-Martin, J.A., Rebolo, R., Carreira, P., et al.
           2003, MNRAS 341, 1084
\bibitem{} Sievers, J.L., Bond, J.R., Cartwright, J.K., et al. 2003,
           ApJ 591, 599  
\bibitem{} Spergel, D.N., Verde, L., Peiris, H.V., et al. 2003, ApJS 148, 175
\bibitem{} Wang, L., Steinhardt, P.J. 1988, ApJ 508, 483
\bibitem{} Wang, L. 2006, ApJ 640, 18
\bibitem{} Weinberg, N.N., Kamionkonsky M. 2003, MNRAS 341, 251
\bibitem{} Wetterich, C. 1988, Nucl. Phys. B 302, 668
\bibitem{} Wetterich, C. 1995, A\&A 301, 321
\end{thebibliography}
\end{document}